\documentclass[twocolumn,showpacs,preprintnumbers,amsmath,amssymb]{revtex4}
\usepackage{graphicx}
\usepackage{amsmath}
\usepackage{titlesec}
\usepackage{feynmp}
\usepackage{float}
\usepackage{epstopdf}
\usepackage{slashed}
\DeclareGraphicsExtensions{.pdf,.png,.jpg}

\titleformat{\section}[runin]{\bfseries}{}{}{}
\begin{document}

\title{Dynamical Gauge Effects and Holographic Scaling of Non-Equilibrium Motion in a Disordered and Dissipative Atomic Gas}
\author{Jianshi Zhao}
\author{Craig Price}
\author{Qi Liu}
\author{Louis Jacome}
\author{Nathan Gemelke}
\address{The Pennsylvania State University \\ State College, PA 16802, USA}

\begin{abstract} We present a table-top realization of a non-equilibrium quantum system described by a dynamical gauge field propagating on an effectively curved space and time manifold. The system is formed by neutral atoms interacting with both a conservative disordered optical field and a dissipative pumping field. In the presence of a sufficiently dark state, we demonstrate non-equilibrium behavior reminiscent of the information paradox in black hole physics.  At a well-defined transition point, the analog of gauge-boson mass is seen to vanish, inducing scale-invariant behavior as a Higgs-like mechanism is removed.  The subsequent scaling behavior can be understood using the holographic principle with a tunable analog of the Planck length derived from the scaling of disorder.  These effects suggest a range of new phenomena in weakly dissipative quantum systems, including the presence of emergent forms of analog gravitation.

\end{abstract}

\maketitle

\begin{figure*}[htp]
\includegraphics[width=7.1 in]{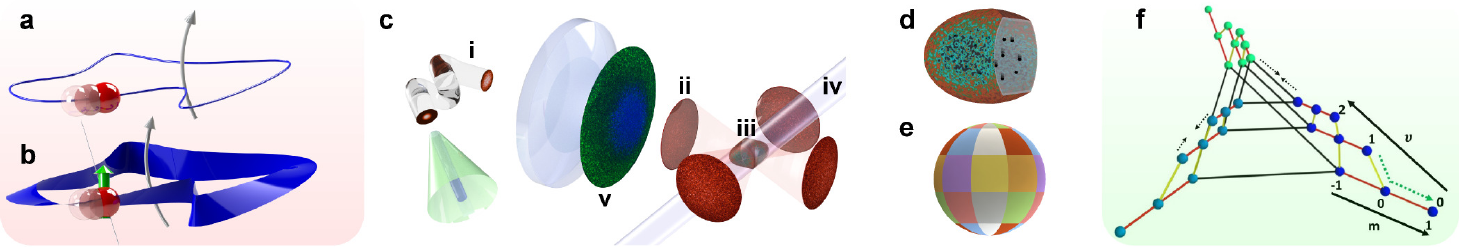}
	\caption{\textbf{Experimental Apparatus.}
    \textbf{a}, Non-equilibrium localization and transport in a disordered system can be understood by interference between time-reversed pathways for particle motion around closed loops, which under time-reversal symmetry coherently interfere, but under dissipative coupling to an external body can be explained by non-vanishing flux (arrow) of an effective gauge field $A^{\mu}_d$.
    \textbf{b}, Here, a neutral atom (red) carrying hyperfine spin (green arrow) propagates in an spin-dependent disordered optical potential, coupled to a dissipative bath which effectively measures the spin along a locally defined axis (normal to the blue plane). In a gauge-orbit $A^{\mu}_d$ defined by projection onto this local axis, transport is described by a classical SU(2) Yang-Mills equation of motion in a full gauge $A^{\mu}_H$.
	\textbf{c}, The microscopy setup.  An equilibrium distribution of optical fiber (i) modes projects four spatially disordered wavefronts (ii), to form a far-off-resonant potential in a small volume (iii), in which atoms are dissipatively coupled to a single-mode optical pumping beam (iv).  Information carried by wavefronts can be controlled by selective launch of optical fibers, yielding altered structure in the aperture plane (v), but unaltered real-space distribution (vi).
	 \textbf{d}, Atomic motion under gravity resembles ``directed percolation", in which atoms are trapped in long-lived metastable states without strong tunneling to neighboring potential minima.
	\textbf{e}, Scaling laws for transport can be understood through scaling of the Shannon-capacity of the optical modes, which counts effective degrees-of-freedom defining $A^{a\mu}_H$ over a coarse-grained surface.
	\textbf{f}, Geometrically, these are points of convexity in a space in which the corresponding master-equation sets the distance between states; such geometry must be defined covariantly through the effective gauge fields $A^{a\mu}_H$ and $A^{a\mu}_d$, which completely specify the dynamics.  Similar ideas used in a continuous sense lead to an analog model for emergent quantum gravity (Methods).
}\label{setupandmodel}
\end{figure*}

At the heart of both gauge symmetry and the second-law of thermodynamics is a redundant labeling of the quantum states of a system.  In the former, this is a consequence of local symmetries, while in the latter it arises from discarding information regarding the state of an external body.  A derivative type of equivalence is exploited in recent work on black-hole physics \cite{Hooft93}, quantum gravity \cite{susskind95} and attacks on long-standing open questions, such as the existence at large length scales of a gauge boson mass gap in Quantum-Chromodynamics (QCD).  This suggests that the study of table-top systems in which the nonequilibrium behavior of a quantized gauge field is sufficiently slow to be observed can shed insight on several problems of broad significance.

To introduce the idea, we first note that for \emph{any} open single-particle system whose basic interaction can be written as $H=\vec{p}^2/2m + V(q) + V_{int}(q,Q)$, a trivial gauge transformation can take the energy into $H'=(\vec{p}-\vec{A}_H(q,Q))^2/2m$, where we denote by $p,q$ the particle momenta and coordinates, $V$ its potential energy, and $V_{int}$ its interaction with its environment with degrees-of-freedom $Q$.  Without further manipulation, the four-vector potential $A^\mu_H=(\phi,\vec{A}_H)$, with $\phi$ the scalar potential, is from a geometric picture trivial in that its field curvature $F^{\mu\nu}=\partial^\mu A_H^\nu - \partial^\nu A_H^\mu$ must vanish everywhere.  However, since the $V_{int}(q,Q)$ must include degrees-of-freedom $Q$ from the environment, any probabilistic description using a density-matrix, $\rho(q,q',t)$, must generally account for ignorance of the state of $Q$ by incorporating nonconservative dynamics into the evolution.  As a result, the curvature $F^{\mu\nu}(q)$ used to predict probabilities for the reduced system no longer vanishes, and $A^\mu_H$ lies on the orbit of a fundamentally irreducible and \emph{dynamical} gauge field degree of freedom $A^\mu_d(q)$ arising from this (potentially dissipative) interaction, without approximation, and despite the \emph{a priori} lack of electromagnetic fields, free charges, or spins in either body.  Furthermore, in a weak coupling limit, the quantization of $A_d$ and $A_H$ become relevant to these dynamics.


While it is not immediately clear such a description is useful, it does suggest that simple geometric aspects underly the statistics of particle motion with dissipation.  For example, even at the classical limit, through a recent generalization~\cite{Feng11} of Berry's geometric phase, one then might expect that bringing the particle along a closed loop $q(t)$ may modify $\rho(q,q')$ by a ``Berry's entropy" $s_B$ related to the curvature of the gauge field~\cite{Feng11} threading the loop, and given for an initially pure state by the final $\rho \ln \rho$, or entropy of entanglement with $Q$.  If $Q$ represent weakly-coupled light-scattering modes for atoms in the presence of a strong conservative potential, the final state \emph{heat energy} $k_BT s_B$ may be seen as a topological redefinition of the photon recoil.

For a thermal wave-packet under continuous weak optical cooling, this can result in Unruh-like ~\cite{Unruh76,Scully03} effects, making the photon and dynamical gauge vacuums experienced by slowly-scattering atoms indistinguishable from a thermal reservoir at a temperature $T_u=\hbar a / 2\pi  k_B v$ with $a,v$ the local wave-packet acceleration and velocity. For atoms with sufficient slow-light-like effects induced by weak coupling fields (Methods), $T_u,a,$ and $v$ can be set on a common scale by the photon recoil.  The resulting motion of atoms is a reflection of the underlying geometric relations of $A^\mu_H$ and $A^\mu_d$ and the quantization properties of the dynamical field $A^\mu_d$.

One finds (see Methods and below) a Higgs-like transition for the gauge-boson mass due to a topological change in a related 2D conformal field theory, and verifiable non-equilibrium scaling laws understood by holographic equipartition familiar from ideas in quantum gravity \cite{PADMANABHAN10}, where the information content of the conservative potential defines the reservior's cooling rate through a number of degrees of freedom $F_s$ defined on ``causal" surfaces. Due to the presence of strong disorder, neither of these effects is simply understood from a perturbative picture and quantum field theory. However, non-perturbative tools, such as those provided by T-, S-, and holographic-duality, and mapping to effectively two-dimensional quantum field theories provide useful alternatives (Methods).  

Here, we investigate non-equilibrium thermal motion of a hyperfine-spin ($F=1$) carrying neutral atom in a disordered environment defined by many interfering far-detuned laser modes $Q$, which introduce a spin-dependent local potential $V_{int}(q,\vec{F},Q)$ \cite{JessenDeutsch} with experimentally quantifiable information content. Additional coupling of \emph{weak} near-resonant pumping modes to the atomic spin results in an additional dissipative spin-interaction with long coherence times.  The gauge potential $A^{a\mu}_d$ induced by discarding information regarding near-resonant modes in the classical limit describes non-equilibrium atomic motion through scale-invariant non-Abelian Yang-Mills equations-of-motion (YM-EOM) \cite{YangMills54,spielmanreview14,Bagchi15} in four dimensions, $\partial^\mu F^{a}_{d\mu\nu} + \hbar g_A \epsilon^{abc}A_H^{b\mu}F^c_{d\mu\nu} = \tilde{j}^a_\nu-j^a_\nu + 2g_\rho\rho^{ab}A^b_{H\nu}$, where $\rho^{ab}$ represents the spin-densities, $j^a$ its currents, $a$ runs over four indices reflecting the U(1) and SU(2) spin symmetry, and the $\tilde{j}^{a}_{\nu}$ is a classically fluctuating (Langevin) source term for the non-vanishing dissipative fields $F_d^{a\mu\nu}$ derived from $A_d^{a\mu}$.

Despite the nonlinearity ($g_A\neq0$), field dynamics exhibit an invariance under a common rescaling of coordinates and time at the classical level, which is lost in a quantized limit. YM theories with such \emph{quantized} gauge fields form portions of the current understanding of quantum-electrodynamics (QED), -chromodynamics (QCD), and the standard model of particle physics, yet ambiguity has remained, in the non-Abelian case of QCD for nearly four decades, concerning existence of gauge boson mass gaps at low (non-relativistic) momenta \cite{Gribov78} and the associated introduction of preferred scales.  This theoretical ambiguity, progressively reversed and resolved by introduction of Faddeev - Popov ghost fields \cite{Faddeev67}, so-called ``BRST-quantization" \cite{Becchi74}, and finally heroic lattice gauge computations \cite{Mendes07}, originates from the possibility of over-counting physically indistinguishable gauges. Here, the ``gauge-" and ``gauge-orbit-" fixing are experimentally tied to the atom-optical interactions through coupling to two separate statistical reservoirs of optical modes, forming an experimental embodiment of stochastic quantization \cite{Nelson66}, and probing the existence of analog gaps by the non-equilibrium evolution of the open system. 

The ensuing YM dynamics can more physically be understood as a geometric effect of measurement back-action, following the suggestion \cite{Aharanov98} that measurement of a spin along an axis defined by a particle's coordinate imparts an effective monopole-like gauge-field. The strong measurement limit, in which the hyperfine spin adiabatically follows the local basis corresponds to an expectation value $_0A_d^{a\mu}$ for $A_d^{a\mu}$ which depends linearly on time (representing a ballistic particle current), and which deviates in the weak measurement limit by introducing an additional dependence $_1A_d^{a\mu}$ independent of time (Methods), which can be interpreted as a dissipative current. The addition of static disorder on its own would be expected to give rise to equilibrium states and transport phenomena exhibiting weak (anti-)localization \cite{Anderson58} modified by spin-orbit coupling, which can be understood by the interference of time-reversed particle pathways (Fig.~\ref{setupandmodel}) along closed loops.  In the presence of \emph{dissipative} flux, the additional breaking of time-reversal symmetry can continuously transfer \emph{information} (and concomitant \emph{heat energy}~\cite{Feng11,Yamano13}), regarding the local disorder to its environment.  We observe under such conditions a non-equilibrium steady-state (NESS) transition and compare its scaling properties to a holographic scaling of the information content in the disordered potential described by $A^\mu_H$.


A static random medium is first introduced using the interference of many randomly-phased optical modes of far off-resonance ($\lambda$=1064 nm, $8$W) laser light, produced by propagation of initially single-mode laser beams through large diameter (910$\mu$m, NA$\sim0.22$) optical fibers to produce an equilibrium distribution of power among their $O(10^5)$ modes. The facet of each fiber is then imaged onto a $500\mu$m diameter volume from four orthogonal directions on the horizontal plane onto a laser-cooled cloud of $^{87}$Rb atoms through high numerical aperture (NA$\sim0.4$) optics (Fig.~\ref{setupandmodel}c). This produces a random network of local potential minima for atoms at the intensity maxima of the laser field. Due to the many interfering modes, the random potential within a small volume $v$ is specified by a quantity (Shannon entropy) of information $S_e$ which scales larger as the volume $v$ is increased.  We have determined through analysis of the fiber mode structure that $e^{S_e} \sim v^{\kappa/3}$, with $\kappa=0.54\pm0.06$ (Methods) under full equilibration.  This disorder varies the locally-preferred dynamics of the gauge-field. Parameterizing disorder in a Kaluza-Klein-like picture with additional coordinates introduced to describe the realization of disorder, an effective gravitational field emerges as described by a local gravitational metric determined by $F_s$.

The experiment begins with $O(10^6)$ atoms transferred into a small (3mm) magneto-optical trap (MOT) in the active region of the optical microscopy setup.  Atoms are first cooled in the presence of the disordered optical field using a far-detuned optical molasses to a temperature of $\sim4\mu$K and then adiabatically released into the random potential, leading to capture of $80\%$ and rapid loss of the remaining atoms, presumably forming the high-energy tail of the thermal distribution. The remaining atoms are allowed to dynamically evolve in the disordered potential for a variable time up to 600 seconds, before being recollected in the MOT to be counted over large dynamic ranges by their fluorescence rate into the microscope.

\begin{figure*}[htp]
\includegraphics[width=7. in]{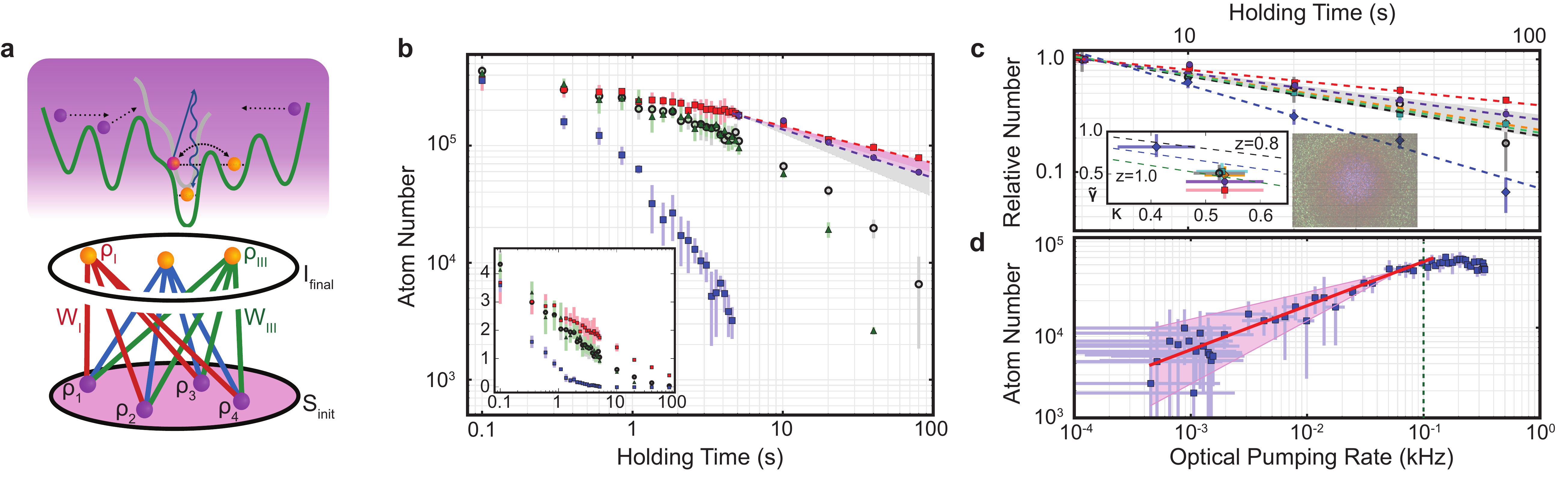}
\caption{\textbf{Dissipation and scaling behavior}
\textbf{a}, The scheme for dissipative coupling, in which atoms in various spin states (orange/red) transition between states and associated potential contours (green/grey) via Larmor precession in an externally defined magnetic and optical field, and are optically pumped (blue) into local vibrational ground states.  Due to the disorder and dissipation, different initial atomic ensembles $\rho_i$ will evolve according to a disorder-specific evolution $w_{j}$ into common local final states $\rho_{j}$, revealing information $I_{\text{final}}$ about the local potential. \textbf{b}, Atoms remaining in the disordered potential as a function of holding time under conservative (green), dissipative (red, purple), non-optimal dissipative (black), and heating conditions (blue).  Dashed lines show power law fits to the dissipative data with optimized cooling for two nominally identical experiment runs $\tilde{\gamma}=0.31\pm0.04$ and $\tilde{\gamma}=0.41\pm0.04$. This data is corrected for vacuum lifetime loss; error bars indicate standard error in the mean due primarily to initial atom number fluctuation. \textbf{b}, inset, Raw data without vacuum loss correction. \textbf{c}, Normalized loss curves for different optical fiber mode content (corresponding to fourier-plane images shown in false color), including equilibrated mode content (red,purple,orange $\kappa=0.54\pm0.06$), low transverse-wavenumber excitation (blue $\kappa=0.41\pm0.07$), and high transverse-wavenumber (black $\kappa=0.52\pm0.05$, green $\kappa=0.53\pm0.05$).  Inset shows loss scaling parameter $\tilde{\gamma}$ against $\kappa$. \textbf{d}, Variation of dissipative capture with pumping intensity. The number of retained atoms after 5.1s is plotted against optical pumping intensity, calibrated to scattering rate from the (bright) $m_F=0$ state. The red line is a power law fit of the data, showing an exponent of $0.5\pm0.2$.  Error bars show the standard deviation in the mean along both axes, and the shaded region uncertainty in the power-law scaling.
\label{holding}}
\end{figure*}

During the evolution time, a controlled dissipation is introduced using an extremely \emph{slow} and broadband form of degenerate Raman sideband cooling (RSC) \cite{Salomon98,Vuletic98} in the disordered optical field (Fig. \ref{holding}a).  Coupling between internal states in the $F=1$ manifold and between vibrational states at the same potential minimum is provided by crossed polarizations of laser light forming the disordered potential, which interact through a vector light shift randomly varying in space. An extremely weak ($3-300$ nW/cm$^2$) optical pumping field, nearly resonant (12MHz blue detuned) with the $|F=1\rangle \rightarrow |F'=0\rangle$ transition on the D$_2$ line provides dissipative coupling quadratic in the hyperfine spin, which allows for a long evolution in the conservative potential before a single spontaneous scattering event.  Details of these interactions, and the resulting YM fields $A_H^{a\mu}$ and $A_d^{a\mu}$, are provided in the methods.  When a magnetic field (57 mG) is applied nearly along the propagation direction of the $\sigma^{+}$-polarized optical pumping beam, a net cooling effect occurs due to the irreversible transfer of energy into spontaneously scattered pumping light, which reverts to heating when the field direction is reversed. On this transition, RSC features nearly-terminal optically dark states due to the tensor-like spin interaction, in which resonance fluorescence cycles cease due to selection rules which decouple the pumping light. However, no single metastable minimum is completely ``dark" to continued motion due to tunneling and the uniform force of gravity. We present (Methods)  microwave-spectroscopic measurements of sideband asymmetry exhibited by atoms after cooling, as well as vibrational spectroscopy measurements, together indicating a kinetic temperature after cooling for short durations below $50nK$, well above the Anderson localization temperature \cite{Inguscio07}, though likely dominated by a thermal tail.  Under optimized conditions, atoms are detectable under continuous cooling for periods as long as 20 minutes.


Following rapid cooling into local motional ground-states, atoms undergo a slow non-equilibrium evolution in the disordered potential under the influence of gravity.  During this time, atoms are lost due to rare (Levy) events in which sufficient kinetic energy is gained that particles undergo unbounded motion, and due to background gas collisions.  In Fig. \ref{holding}, we show the measured atom loss, both as measured, and corrected for vacuum lifetime, exhibiting under heating, no cooling, and non-optimal cooling conditions exponential loss of atom number.  For optimized cooling, power-law decay is observed, with number $N(t) \sim t^{-\tilde{\gamma}}$, with $t$ the cooling time, and $\tilde{\gamma}=0.40 \pm 0.07$.


Anomalous transport \cite{LevyStat,Aspect94} or presence of a non-equilibrium critical point could explain this scale invariance, the latter suggesting connection to directed percolation (DP)\cite{Grassberger79} and absorbing-state phase transitions (ASPT) \cite{munoz97}.  A discussion of the latter is provided in the Methods, but suffers from two shortcomings - (I) ASPT noise-correlations miss topologically-induced (Langevin) noise (as captured by YM), leading to an inapplicable definition of upper-critical dimension. (II) Necessary modifications of critical point phenomena due to strong disorder through the Harris \cite{Harris74} and Luck \cite{Luck93} criteria provide no clear remedy for obtaining scaling exponents.  Similarly, it is unclear how to make simple extensions of Anderson and weak localization phenomena to the present dissipative case in a way to predict scaling laws.

Instead, we first provide a simple argument, independent of the gauge description, spatial metric, and YM that power-law decay and the role of disorder can be understood more simply geometrically, and from an approximate discretized model more familiar in laser-cooling methods. On long timescales, the diagonal components of a density matrix $\hat{\rho}_a(t+\Delta t)=\sum_{Q',Q} \langle Q' |\hat{u}^\dagger(\Delta t)|Q \rangle  \hat{\rho}_a(t)\langle Q|\hat{u}(\Delta t)|Q'\rangle$ should suffice to describe transport between discretized quantum states.  The evolution is Liouvillian due to the projective action $|Q\rangle\langle Q|$ tracing over optical states $|Q\rangle$; in general the master equation matrix element $w_{ij}=\partial \rho_{a,ii}(t+\Delta t)/\partial \rho_{a,jj}(t)\neq w_{ji}$. For long measurement times, the loss time constant reflects the smallest relevant eigenvalue of the operator $\omega_{ij} = \lim_{\Delta t\rightarrow 0} (w_{ij}-\delta_{ij})/\Delta t$; power-law behavior results when this eigenvalue vanishes in the long-time limit as $\tilde{\gamma}/t$, and indicates an absence of dynamic scale.  The change in phase-volume in a single time-step is given by $\det(w)$ - its second largest eigenvalue $\lambda_2$ (representing the slowest process away from equilibrium) is bounded by Cheeger's- \cite{Chung05} and Bruger's- inequalities through the isoperimetric constant $h$ as $1 - h^2/2 \geq \lambda_2 \geq 1 - 2h $ - the vanishing constant $\gamma$ implies sub-exponential decay $h\sim t^{-\gamma_h}$, with $1 \leq\gamma_h\leq 2$. The isoperimetric constant represents the smallest surface-area to volume ratio of a discrete space (Fig. \ref{setupandmodel}f) of atomic states describing directed transport, and suggests a similar underlying continuous geometric model without discretization.

\begin{figure}[htp] \label{higgs}
	\includegraphics[width=3.5 in]{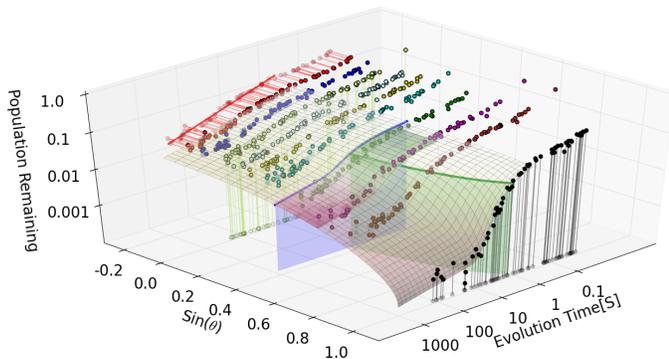}
	\caption{\textbf{Higgs-like transition to scaling behavior.} The number of atoms retained in the disordered potential (corrected for vacuum lifetime) after varied holding times and for varied orientation of magnetic and optical pumping fields are shown as scatter points. The transition to scaling-law behavior (blue line, $\theta=23\pm2^o$) is sharp, representing a discrete transition between decay due to massive (exponential, time-constant shown by green line) and massless (scale-invariant) gauge boson modes. Here, $\theta$ represents the angle between the external magnetic field and optical pumping beam propagation axis, which determines degree of decoupling of the dark state. Zero angle corresponds to pure $\sigma_+$ polarization, which maximally decouples the dark state.  The mesh surface shows a fit result from a model allowing for a crossover from exponential to power law decay of varying width. }\label{higgs}
\end{figure}

\begin{figure}[htp] \label{rscooling}
	\includegraphics[width=3.25 in]{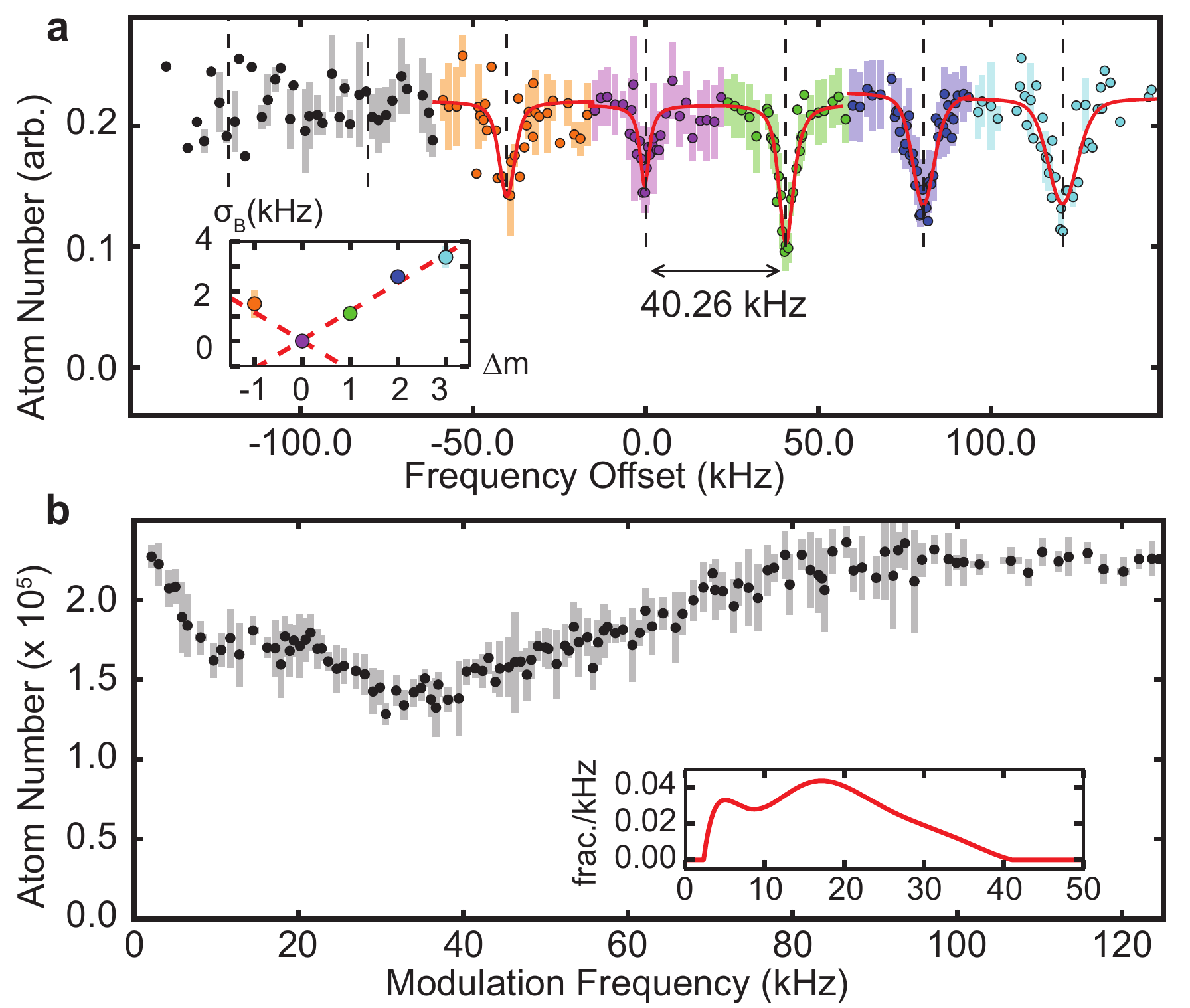}
\caption{\textbf{Characterization of the disordered potential.}
	 \textbf{a}  Microwave spectroscopy showing sideband asymmetry of the atoms after cooling.
	The asymmetry of each peak (see inset) can be used to extract kinetic temperature - the red line shows a fit (Methods) to parameterize temperature. \textbf{b} Parametric excitation of the atoms is performed by modulating the light intensity; this couples vibrational levels that differ by two due to even symmetry and is used (inset) to infer the distribution of local trapping frequencies. The plot shows the survived atom number as a function of the center frequency of the modulation. The inset shows the fraction of atoms contained within a 1kHz bin of local vibration frequency. }\label{rscooling}
\end{figure}

This geometric bottle-necking can be understood on physical grounds as a non-equilibrium quantum effect. The thermal length scale for the gauge-field dynamics is set by the thermal deBroglie wavelength $\lambda_{db}$ for a cooling atomic wavepacket.  Within $\lambda_{db}$, excitations of the gauge-field should be related by unitary dynamics, whereas points separated by more than $\lambda_{db}$ are related predominantly through classical statistical correlations.  The boundary between these regions, including time as a coordinate, defines a type of horizon similar to that considered in black-hole physics, where the holographic principle was first discussed.  This boundary becomes sharp if, as there, the speed of propagation (group velocity) for unitary dynamics vanishes.  In both systems, this boundary must accelerate at some rate $a$; the resonant vacuum modes in the atomic system then appear thermally occupied at an Unruh temperature $T_u=(\hbar a/ 2\pi c k_B)$, and a balance is achieved in the free energy changes due to thermal entropy and the entropy of disorder. The relation to gravity is not just cosmetic - through detailed modeling of this behavior (Methods), one may arrive at a Palatini formulation for general relativity in a set of non-commutative coordinates describing optical modes and excitations of the gauge field. The Higgs mechanism can be understood by mapping the dynamics of the gauge-field onto a related two-dimensional conformal field theory, in which the dark-state cooling conditions appear effectively as a gauge potential acting on the dynamics of a two-dimensional sheet.  This gives mass to the gauge boson with sufficiently compromised dark-state cooling, and exhibits a sudden transition to masslessness with sufficiently dark terminal states, corresponding to vanishing flux on the sheet. This transition is observable by varying the angle subtending the optical pumping propagation axis and the external magnetic field (Fig. \ref{higgs}).

In the massless case, the Cardy formula \cite{Cardy} can be used to relate the variation of thermal entropy and that of disorder, showing a type of dynamic equilibriation which relates the associated disorder and thermal temperatures to a local gravitational field strength (a dilaton field) - under such conditions, a holographic scaling law is a natural outgrowth, relating scaling powers for degrees of freedom in the optical potential and the dynamic evolution time.  In fig. \ref{holding}, we demonstrate such an effect, observing a relation between the atom loss constant $\tilde{\gamma}$ and the information scaling $\kappa$ roughly consistent with $\kappa-2 = -z(\tilde{\gamma}+1)$, with a dynamic critical exponent $z=1$, suggesting the relevant scaling quantity is the informational degrees of freedom per unit area.

The scale invariance of the classical YM dynamics should be lost in the quantum limit, as the coupling constant $g_A$ now introduces a preferred scale for the dynamic evolution.  To explore this regime, the experiment was repeated at a series of optical pumping intensities for fixed experiment duration, see figure \ref{holding}d.  The scaling of retained number for sufficiently low intensities follows an apparent power-law form $N_v\sim I^{\lambda}_{op}$, with $\lambda=0.5\pm0.2$ over two decades in intensity $I_{op}$.  While not fully understood, we note that the number of atoms retained follows roughly the same dependence as the dissipative flux density $\vec{\nabla} \times \vec{A}_d$, which suggests the relevance of a non-equilibrium version of ``filling-factor." This is consistent with the picture above, in which the dissipative flux associated with the dynamical gauge-field sets a relevant scale for disorder.

In conclusion, we have demonstrated an experimental platform in which dissipative non-equilibrium effects in a thermal sample can be directly tied to microscopic parameters through an effective gauge-field formalism following the YM equations of motion in a curved background metric driven by disorder.  A Higgs-like transition is observed in the gauge boson mass, resulting in scale-invariant behavior on the massless side of the transition, which can be directly linked through holographic principles to the measures of information in the underlying disordered potential.  We have explored the non-classical limit of the same model, and observe a different scale-invariant behavior, which we tentatively associate with a non-equilibrium version of flux-binding.  The utility of holographic scaling arguments and existence of models built on ideas in quantum gravity suggest a wide array of new phenomena in weakly dissipative quantum systems with disorder.  This work also points to new, highly feasible routes to the study of topologically ordered systems, by exploiting the non-equilibrium response of cold-atom systems rather than their low-temperature equilibrium properties, and highlights the utility of geometric and topological approaches to the study of systems with strong disorder.

\newpage
\newpage

\section*{Methods}

\subsection{Dissipative Gauge Potential}

The dissipative optical potential used in this experiment is somewhat novel in several regards - it is \emph{weakly} dissipative, disordered, and includes dark-state cooling effects.  While not necessary, it is convenient to describe the atomic motion in its totality using a single gauge potential.  To capture dissipative effects at the single-photon scattering level, it is necessary to incorporate entropic effects into the description of the gauge field.  Below, we will incorporate this naturally into the gauge-freedom already present in the description.  To capture the quantum effects associated with slow scattering from a multilevel atom, we will show that it is necessary to include Unruh-like effects, which are naturally captured by using geometric principles in the construction of normal modes for optical scattering.  Finally, to incorporate disorder, it is useful to draw on concepts from quantum gravity and particle physics, which are captured naturally by the gauge field description.

To begin, we take the atom-optical interaction and gravitational potential to have a Hamiltonian of the form

\begin{equation}
H = \vec{p}^2/2m + U_0 \vec{E}_f\,^\dagger\cdot\vec{E}_f + (g \vec{\mathcal{B}}+d \vec{\mathcal{D}}) \cdot \vec{F} + mgz,
\end{equation}

\noindent with $\vec{p}$ the momentum, $m$ the mass, $z$ the vertical coordinate, and $\vec{F}$ the atomic hyperfine spin operator, $U_0=\hbar\Gamma^2 I / 12 \delta I_s$ the scalar potential with $\Gamma$ the excited state line-width, $\delta=(1/2\delta_{1/2}+1/\delta_{3/2})^{-1}$, $I_s$ the saturation intensity, $g_F$ the Land\'{e} g-factor, and $\vec{E}_f(\vec{x})$ the far-detuned optical field.  The total effective magnetic field

\begin{equation}
\vec{\mathcal{B}}= \vec{B}_{ext} + B_0  \vec{E}_f\,^\dagger\times\vec{E}_f
\end{equation}

\noindent includes the externally applied field $\vec{B}_{ext}$ and the vector light shift characterized by $B_0=i D U_0 /2\hbar$, with the relative detuning $D=(\delta_{3/2}-\delta_{1/2})/(\delta_{3/2}/2+\delta_{1/2})$ of the far-detuned light from the $5^2S_{1/2}$ to $5^2P_{3/2}$ and $5^2P_{1/2}$ transitions. We assume the electronic excited state population due to the optical pumping process coupling the $F=1$ and $F'=0$ is small, and with the excited state adiabatically eliminated, dissipation enters through

\begin{equation}
\vec{\mathcal{D}}=(\vec{F}\cdot \vec{E}_{op})\sum_v \vec{E}^\dagger_v + (\vec{F}\cdot\sum_v \vec{E}_v) \vec{E}^\dagger_{op}.
\end{equation}


\noindent Here, $\vec{E}_{op}(\vec{x}), \vec{E}_v(\vec{x},t)$ are the field operators for the coherent optical pumping and vacuum fields in a rotating-wave picture.

Under a local gauge transform of the hyperfine spin reference direction $\hat{n}_F$ to coincide with that of the local total effective magnetic field $g\vec{\mathcal{B}}+d\vec{\mathcal{D}}$, the atomic wave-function is transformed by $U_1=\exp{(-i F_y \theta_F/\hbar)}\exp{(-i F_z \phi_F/\hbar)}$ with $\theta_F$ and $\phi_F$ orientation angles for $\hat{n}_F$.  As a result, the transformed interaction $H'=U_1^\dagger H U_1$ acquires a non-abelian gauge field $A^{a\mu}$ through $p^i\rightarrow p^i - g_A F^a A^{ai} $, with $g_A$ introduced as a dimensionless coupling constant to understand subsequent scaling transformations.  The interaction between the optical field and atoms becomes $V_{int} = Q F^x$ with $Q(\vec{F})=|g \vec{\mathcal{B}}+d \vec{\mathcal{D}}|$ effective degrees-of-freedom of the `classical bodies' formed by the optical fields, which interact with the transformed spin component $F^x$ either directly, or via a process second order in the spin. The gauge field $A^{a\mu}$ varies in time and space due to the direct spatial and temporal-dependence of the vacuum or far-detuned modes $\vec{E}_v(\vec{x},t),\vec{E}_f(\vec{x})$, but without further manipulation forms a pure (or trivial) gauge field in that the first two terms of the field curvature

\begin{equation}
F^{a\mu\nu}=\partial^\mu A^{a\nu} - \partial^\nu A^{a\mu} +  g_A \epsilon^{abc} A^{b\mu} A^{c\nu}
\end{equation}

\noindent vanish everywhere identically.

Below we will provide a more careful analysis of the effect of open coupling to the external optical fields, but we first give a more physical feel for its origin by considering the effect of measurement back-action due to the effective measurement formed by coupling to the classical degree-of-freedom $Q$. Under strong coupling of the spin to the optical fields, the spin may be expected to adiabatically follow the locally defined spin basis $F^x$.  As a result of the commutation $[F^a,F^b]=i\hbar\epsilon^{abc}F^c$, the components $F^y,F^z$ exhibit maximal uncertainty, and any motion of the particle according to the influence $A^{a\mu}$ occurs as a coherent summation over all possible pathways of $F^y,F^z$; the rapid dephasing over neighboring pathways effectively fix the $F^y,F^z$ contributions to zero, and the corresponding field curvature no longer vanishes.  To better clarify the latter effect, through the course of interaction, the value of the optical degree of freedom $\pi_Q$ conjugate to $Q$ forms an indicator of the time-averaged spin component $F^x$ of the atom in the local basis, and the non-commuting $F^y,F^z$ experience measurement back-action.

It is helpful for further calculation to introduce an alternative gauge formulation through an added transform $U_2=\exp{(i f_2)}$ with $f_2 = - t (Q F^x + V)/\hbar$, with $V$ the remaining scalar potential, such that the atom-optical hamiltonian $H_2=U_1^\dagger U_2^\dagger H U_2 U_1 = (p^i+g_A F^a A^{ai}_H )^2/2m$, with $A^{ai}_H=A^{ai}+A_2^{ai}$, and $A_2^{ai}=\partial^i f_2 $.

It is important to stress that the presence of optical the mode operators present in the definition of $A^{ai}_H$ imply that it is a quantum, or dynamical field, which evolves as light is scattered. Below we will carefully derive the dynamics of this field starting from the Hamiltonian above.  We first point-out that its motion might \emph{ab initio} be expected to follow a generic formulation for a non-abelian gauge field coupled to a spinful atom.  Below we will find that this is not exactly the case, and differs in several important, and measurable, ways.  Nevertheless, the evolution of atomic density might be understood to arise from the non-relativistic limit of a Lagrangian density

\begin{equation}
\mathcal{L}=\mathcal{L}_A+\mathcal{L}_\psi
\end{equation}

\noindent corresponding to a complex three-component scalar field $\psi^a$ representing the atomic state in hyperfine state $a$, and coupled to a SU(2) gauge field $A^{a\mu}$.  The field-dependent part includes gauge-fixing and matter-gauge interactions through

\begin{eqnarray}
\mathcal{L}_A= -F^{a\mu\nu}F^a_{\mu\nu}/2 &+& j^a_\mu A_H^{a\mu} +   g_\rho A^{a}_{H\mu} \rho^{ab} A^{b}_{H\mu} \label{eq::La}
\end{eqnarray}

\noindent with $j^a_\mu=ig_A\partial_\mu \psi^{b*}F^a\psi^b/2mc $ the four-current for density, $\rho^{ab}=(\psi^{a*}\psi^{b} +  i \epsilon^{abc}\rho_F^c)$ a matrix whose diagonal components $\rho^{aa}$ give the density in the spin state $a$, and off-diagonal components $\rho^c_F = \psi^{*a} F^c_{ab} \psi^b/\hbar$ form measures of the local density of spin component $c$. The constant $g_\rho=g_A^2\hbar/2mc$ carries dimensions of length. The field-independent portion is

\begin{equation}
\mathcal{L}_\psi = \frac{\hbar}{2mc}[\partial_\mu \psi^{a*} \partial^\mu \psi^a - \frac{m^2c^2}{\hbar^2} \psi^{a*}\psi^a]
\end{equation}

We have first used the form of a (not rigorously physical) relativistic field theory here to connect more closely to known constructions for the Yang-Mills fields, in which the matter-gauge interaction can be understood from the the square of a covariant derivative $D_\mu=\partial_\mu+i g_A F^a A^a_\mu /\hbar $ acting on $\psi$, though we will in all cases be interested in the non-relativistic limit of the complete coupled system of fields and particles. A more satisfying construct here might be built by taking careful non-relativistic limits of the electroweak/Higgs~\cite{Higgs64}, St\"{u}ckelberg~\cite{Stueckelberg381,Stueckelberg382,Stueckelberg383} or Proca actions~\cite{Proca36}, but we will forgo discussion of these.  To understand the non-relativistic constraints from the model above, we consider the transformation $\psi^a\rightarrow\psi^a \exp{(-imc^2t/\hbar)}$, which can be understood to offset the scalar potential $A^0_0$ by the rest energy.  Discarding terms which go like $(\hbar/mc^2)\partial_t \psi$ has the effect of returning atomic dynamics to a non-relativistic form and eliminating non-physical negative norms and non-conservation of current.  Note that in the following we \emph{will not} make use of relativistic horizons or other special-relativistic phenomena predicated on the value of the speed of light itself.

The probability amplitude for traversal of the atomic wave-function $\psi^a$ along a path is then, with a fully specified $A^{a\mu}$, given in a path-integral form as $W[\psi^a | A^{a\mu}]=\exp{(i S[\psi^a | A^{a\mu}])}$, with $S=\int dt d^3x \, \mathcal{L}$.  To obtain full atomic dynamics without regard to the optical state, the probability $W[A^{a\mu}]$ for a given trajectory of $A^{a\mu}$ must be appended, and integrated over.  

Unfortunately, the $O(N)$ scalar field theory written above to model $F=1$ atoms through $\psi^a$ does not follow proper symmetries under relativistic coordinate transformations, and suffers complication (anomalies) particularly after quantization.  Therefore, instead of taking this traditional route of writing down a classical lagrangian like the one above, and attempting to quantize it in ways to avoid anomalies, we take advantage of our well-understood quantum system, and develop a gauge-formalism from the bottom-up. This preserves anomaly-free behavior as the dynamical gauge-field picture is built.


The gauge field is specified through the transformation $U=U_2U_1$ by

\begin{eqnarray}
\bar{A}_\mu = i[\partial_\mu U]U^{-1} = i\partial_\mu \log U.
\end{eqnarray}

\noindent Locally, this is a member of the Lie group for rotations in SU(2) generated by the $F^a$ and extended by U(1) spin-independent transformations, and will later be expressed as

\begin{eqnarray}
\bar{A}^\mu = \sum_{b=0...3}  \bar{A}^{b\mu} F^b /\hbar
\end{eqnarray}

\noindent where $\bar{A}^{b\mu}$ is a function of the coordinates and the optical fields, and the SU(2) generators have been extended by one index $b=0$ such that $F^{b=0}$ represents $\hbar$ multiplying the hyperfine spin identity operator.  The lie-group algebra must then be extended such that $[F^a,F^b]=i\hbar\epsilon^{abc}F^c$ with the structure constant $\epsilon^{abc}$ now running over four indices, and $\epsilon^{abc}=0$ for any $a,b,c=0$.  We likewise extend $\rho^{ab}$ such that $\rho^{00}=\sum_a |\psi^a|^2$, and $\rho^{0a}=\rho^{a0}=0$.  This provides a convenient form for expressing both the gauge-fields and particle- and spin-currents, and $j^a_\mu$ can follow a similar extension, which incorporates the total mass flow in the $a=0$ component.

We now ``fix" the gauge field explicitly to this form using a path-integration

\begin{eqnarray}
W[A^{b\mu}]=&\int& D[\{a\}] \prod_b \delta[A^{b\mu} - \bar{A}^{b\mu}] \nonumber\\ &\prod_i &\delta[\vec{E}_i-\sum_j a^j_i \vec{E}^j_i(\vec{x},t)] W[\{a\}],
\end{eqnarray}

\noindent where we express the electric fields in terms of individual mode amplitudes $a$, which are taken to represent coherent state amplitudes, and the action in $W[\{a\} ]$ is given by the $\mathcal{S}_a=\int dt (\dot{a_i}^{j*}a^j_i - \omega^j_i a^{j*}_ia^j_i) $ where a sum is implied over all three mode types $i$ and individual modes $j$ within each class. For the far-detuned modes $E_{fi}$ and optical pumping mode $E_{op}$ we set $\omega^j_{i=\epsilon,op}=0$, but for the vacuum modes the frequency of $E_\nu$ is measured relative to the coherent optical pumping mode frequency.  Expressing the delta function above through a fourier-transform, and absorbing a constant,

\begin{eqnarray}
W[ A^{b\mu} ]  =  \int  &D[\{a\},\tilde{j}^b_\mu ]\exp{[i\int dt (\dot{a_i}^{j*}a^j_i - \omega^j_i a^{j*}_ia^j_i)]} \nonumber \\
&\times\exp{(i\int d^3x \,dt[\tilde{j}^b_\mu(A^{b\mu}   -  \bar{A}^{b\mu})])}
\end{eqnarray}

\noindent we have, using eq. \ref{eq::La}, an Euler-Lagrange equation of motion from variation of $A^b_\mu$ as

\begin{eqnarray}
\partial^\mu F^{a}_{\mu\nu} &+& \hbar g_A \epsilon^{abc}A^{b\mu}F^c_{\mu\nu}  \\ &=& \tilde{j}^a_\nu-j^a_\nu - 2 g_\rho\rho^{ab} A^b_\nu  \nonumber
\end{eqnarray}

\noindent These dynamics are not complete in the sense that they depend on as yet unspecified dynamics of the mode amplitudes $\{a_i\}$ through $\tilde{j}^{b\mu}$.  Rather than develop a second equation of motion, which would result in full deterministic evolution of the coupled atom-optical system (with a vanishing field curvature as a consequence of the ``pure gauge" represented by $\bar{A}^b_\mu$), we integrate-out the modes $\{a_v\}$ corresponding to the vacuum modes of the pumping process.  We separate the pumping-dependent terms from the gauge term $\bar{A}^b_\mu$, such that the pumping-dependent portion of the Lagrangian density becomes

\begin{eqnarray}
\mathcal{L}_D \approx \tilde{j}^b_\mu\bar{A}^b_{D\mu}[\{a_\nu\}] + \sum_j |E_\nu^j|^2 a^j_\nu(i\partial_t - \omega_\nu^j) a^{j*}_\nu / c,
\end{eqnarray}

\noindent where $\bar{A}^b_{D\mu}$ is the first order correction to $\bar{A}^b_{\mu}$ at low coherent optical pumping mode strength $a_{op}$. It is helpful to first define

%

\begin{eqnarray}
\bar{A}_{D\mu} = a_{op}\lim_{a_{op}\rightarrow 0} \frac{\partial \bar{A}_{\mu}}{\partial a_{op}}  = i a_{op} [\partial_\mu \partial_{a_{op}} \log U ]_{a_{op}=0}
\end{eqnarray}

\noindent and split the rotation $U_1=U_1^\delta U_1^B$ into two parts, first rotating the hyperfine spin axis locally into the direction of $\vec{\mathcal{B}}$, and subsequently into the direction $\hat{n}_F$ through $U_1^\delta$.  A similar split for $U_2$ leads to a simpler expression for $\bar{A}_{D\mu}$ as

\begin{eqnarray}
\bar{A}_{D\mu} = a_{op} \partial_\mu [\partial_{a_{op}} U_2^{\delta} U_2^B U_1^{\delta} (U_2^{B})^{-1}]_{a_{op}=0}
\end{eqnarray}

\noindent which reduces to

\begin{eqnarray}
\bar{A}_{D\mu} = \partial_\mu[-\frac{2td}{\hbar}(\vec{\mathcal{D}}\cdot\hat{\mathcal{B}})(\vec{F}\cdot\hat{\mathcal{B}})-\tilde{\Omega} ]
\end{eqnarray}

\noindent where $\tilde{\Omega}$ represents the effect of measurement back-action on the non-commuting components $F^{y,z}$ in an ``interaction-representation," due to the local differential rotation $\delta \theta_F, \delta \phi_F,$ of the hyperfine spin axis away from the direction specified by $\vec{\mathcal{B}}$ due to $\vec{\mathcal{D}}$.  Explicitly,

\begin{eqnarray}
\tilde{\Omega} = e^{-ig\vec{\mathcal{B}}\cdot\vec{F} t/\hbar} e^{2i(F^y \delta \theta_F + F^z \delta \phi_F)/\hbar} e^{ig\vec{\mathcal{B}}\cdot\vec{F} t/\hbar} -1 \\
\equiv e^{-ig\vec{\mathcal{B}}\cdot\vec{F} t/\hbar} e^{2i\vec{F}\cdot\delta\vec{\theta}_F/\hbar} e^{ig\vec{\mathcal{B}}\cdot\vec{F} t/\hbar} -1 \nonumber \\
= e^{2i(\delta\vec{\theta}_F+gt \vec{\mathcal{B}}\times\delta\vec{\theta}_F)\cdot \vec{F} /\hbar}-1 \nonumber
\end{eqnarray}

\noindent and

\begin{eqnarray}
\bar{A}_{D\mu} = \frac{2d}{\hbar}\vec{F}\cdot \partial_\mu[(\hat{\mathcal{B}}\times\vec{\mathcal{D}}&-&\hat{\mathcal{B}}\cdot \vec{\mathcal{D}}\hat{\mathcal{B}})t \\
&-&\frac{\vec{\mathcal{D}}-\hat{\mathcal{B}}\cdot \vec{\mathcal{D}}\hat{\mathcal{B}}}{g\mathcal{B}}\nonumber
]
\end{eqnarray}

\noindent Due to the SU(2) and U(1) symmetries, the local gauge transformation may be written using the same ``extended" index $b=0...3$, with $F^0/\hbar$ representing the hyperfine spin identity operator, and the remaining three indices the SU(2) generators $F^{x,y,z}$ in a form

\begin{eqnarray}
\bar{A}_{D\mu} = \sum_{b=0...3} \bar{A}^b_{D\mu} F^b /\hbar
\end{eqnarray}


\noindent This can be written concisely as

\begin{eqnarray}
\bar{A}^c_{D\mu} =  \partial_\mu [G^c_{rk} \Sigma_{op}^{rk}]
\end{eqnarray}

\noindent where $\Sigma_{op}^{rk}=\sum_j \Sigma_{op,j}^{rk} =\sum_j (a_{v}^{j*}E^r_{op} E_{v,j}^{\dagger k} +  a_{v}^{j} E_{v,j}^{k} E^{\dagger r}_{op}) $ is the spatial components of the stress-energy tensor for the optical pumping field, and

\begin{eqnarray}
G^c_{rk}  &=& \frac{2d\hbar}{g\mathcal{B}} (\epsilon^{krc}+\delta_{kr}\delta_{c0}) \\ &-& 2d\hbar (\epsilon^{irc}+\delta_{ir}\delta_{c0})(t \epsilon^{ijk}\hat{\mathcal{B}}^j + (\frac{1}{g\mathcal{B}}-t)\hat{\mathcal{B}}^i\hat{\mathcal{B}}^k) \nonumber \\
&\equiv& _0G^c_{rk}  +\, _1G^c_{rk} t  \nonumber
\end{eqnarray}

\noindent The field $\bar{A}^b_{D\mu}$ depends linearly on the vacuum mode amplitudes $a^j_\nu(t)$, and the path integration can be written as a gaussian in $a^j_\nu$.  To understand the contributions from $\tilde{j}^b_\mu$, it is convenient to first integrate the action by parts, obtaining

\begin{eqnarray}
\mathcal{S}_D = \int d^4x \, G^b_{rk} \Sigma_{op}^{rk} (\delta_{S_\mu}-\partial_\mu)\tilde{j}^{b\mu} \\ \nonumber
+ \sum_j |E_\nu^j|^2 a^j_\nu(i\partial_t - \omega_\nu^j) a^{j*}_\nu / c \label{eq::Sd}
\end{eqnarray}

\noindent where $\delta_{S_\mu}$ is the $\mu^{\text{th}}$ component of the unit normal at the bounding surface of the integration.

The quantity $\delta\tilde{\rho}^b=(\delta_{S_\mu}-\partial_\mu)\tilde{j}^{b\mu}$ can be interpreted within the bulk as the deviation of the Langevin current $\tilde{j}^{b}_\mu$ from local conservation (and on the bounding surface $S$ as a flux).  We intend to create a description of the dynamics of this system which does not include detailed information regarding the scattered optical pumping light - integrating out the dynamics of the mode amplitudes $a_\nu^j$ achieves this, and gives an alternate description using a condensed set of physical parameters, including a dynamical gauge field and its evolution according to a modified view of the underlying coordinate system. We implement this by reformulating the field in terms of a local `Bogoliubov-Fourier' transform

\begin{eqnarray}
 a_\nu^j(t) &=& \int d^4\eta d^4x' \,  E_{v,j}^{k*}(x')\delta(t-t') u^k_i\beta_i(\eta_\mu) \nonumber
\end{eqnarray}

\noindent where the $u^k_i[\eta_\mu, x_\mu]$ with $i=(0,1)$ represent spatial `wave-functions' of generalized momenta $\eta_\mu$, we take $\beta_1=\beta_0^*$ to represent anomalous coupling of the vacuum annihilation operator $a^v_j$ to creation operators $\beta_1[\eta_\mu]$, and will later perform path-integration over the new amplitudes $\beta_i$.

In order to account for Unruh-like effects using the mode structures $u^k_j$ (defined explicitly below), in which unoccupied modes appear occupied in an accelerated coordinate system, we choose the dependence on the coordinates to enter through a new, smooth and conformally-mapped, complex coordinate system $\bar{x}_{\mu}[x_\nu]$ - we choose this in the spirit of understanding Unruh effects through analytic continuation of the action of the hamiltonian into Lorentz boosts, and will take the complex part of the new coordinates to generate anomalous couplings using a specific form of $u_j^k[\eta_\mu,x_\mu]$ below.

We further choose the corresponding metric tensor $h^i_{\mu\nu}$ (with $i=0,1$ representing the metric and its conjugate) to satisfy $(\det{h^i})^{1/2}=\bar{G}_i^b\delta\tilde{\rho}^b$, where $\bar{G}_0^b=G^b_{rk}(E^{r*}_{op}u_1^k+E^{r}_{op}u_2^{k*})$ and $\bar{G}_1^b=\bar{G}_0^{b*}$.

The quantity $\bar{G}_1^b$ then represents a sum of `cooling' and anomalous absorbtion processes, in which either a pumping photon is absorbed and re-emitted into an unoccupied vacuum mode, or a photon is absorbed from the apparently thermally occupied vacuum mode by an accelerating atom and emitted into the pumping beam. The quantity $\bar{G}_0^b$ represents the unitary opposite of these processes.  With this definition, the dissipative part of the action can then be written in a simple, more geometric form

\begin{eqnarray}
\mathcal{S}_D &=& \int d^4\eta\ [ v_i \beta_i +  \beta_i\, \kappa_{ij}[\eta]\, \beta_j ]  \nonumber
\end{eqnarray}


\noindent with $v_i=\int d^4\bar{x}_i=\int d^4 x \sqrt{-h^i}\exp{(\phi^i)}$ the volume in the new coordinates $\bar{x}_i$, and under a condition

\begin{eqnarray}
\int d^4{x}\, \sigma^x_{ij}[ u^{k}_l(\eta_1 x) \partial_{t}  u_i^{k*}(\eta_2 x) - u_i^{k*}(\eta_2 x) \partial_{t} u^{k}_l(\eta_1 x)]    \nonumber \\
\equiv -i\kappa_{lj}[\eta_1]\delta(\eta_1-\eta_2)  \label{eq::condition}
\end{eqnarray}

\noindent for some function $\kappa$, which ensures the ``diagonality" in $\eta$, and permits the $\beta$ to be integrated out of the path integral (conditions under which this is possible are discussed below).  This leads to an effective action


\begin{eqnarray}
\mathcal{S}^{\text{eff}}_D =
\int d^4{\eta}\, (v_i \kappa_{ij}^{-1} \, v_j - \frac{1}{2}\bar{c}_i\kappa_{ij}c_j)
\end{eqnarray}

\noindent where we have included the change-in-measure associated with the determinant of the operator $\kappa$ through anti-commuting Grassmann numbers $c_i[\eta]$ and $\bar{c}_i[\eta]$, representing so-called `ghost' modes - while these are decoupled from the dynamics at the classical level, we include them in order to retain the well-behaved quantized path integration we began with - the function $\kappa_{ij}[\eta]$ can now be recognized as the propagator for ghost modes with momenta and energy $\eta_\mu$ in a Yang-Mills picture. The inverse of this operator also relates the dynamical gauge field to the Langevin current. Restoring the original coordinates in the integration,

\begin{eqnarray}
\mathcal{S}^{\text{eff}}_D &=& \int  d^4{x}_1\, d^4{x}_2 \, \delta\tilde{\rho}^b[x_1] V^{bc}_{\tilde{\rho}}[x_1,x_2] \delta\tilde{\rho}^c[x_2]  \nonumber
\end{eqnarray}

\noindent with an effective interaction

\begin{eqnarray}
V^{bc}_{\tilde{\rho}}[x_1,x_2] &=& \int d^4\eta\, \bar{G}_i^b[x_1,\eta]\, \kappa^{-1}_{is}\bar{G}_s^c[x_2,\eta]\label{eq::eff_int}
\end{eqnarray}

\noindent returning to $\tilde{j}_\mu$, and integrating again by parts,

\begin{eqnarray}
\mathcal{S}^{\text{eff}}_D &=& \int  d^4{x}_1\, d^4{x}_2 \,  (\partial^{(1)}_\mu \partial^{(2)}_\nu V^{bc}_{\tilde{\rho}}[x_1,x_2]) \tilde{j}^b_\mu[x_1] \tilde{j}^c_\nu[x_2]  \nonumber \\
&& \equiv \int  d^4{x}\ \tilde{j}^{b}_\mu A_{\tilde{j}}^{b\mu} \label{eq::Sd_v}
\end{eqnarray}

\noindent where one can view the dynamical gauge field $A_{\tilde{j}}$ generated by the Langevin current $\tilde{j}$ as

\begin{eqnarray}
A_{\tilde{j}}^{b\mu}[x] &=& \int d^4{x}_s \,  (\partial^{(x)}_\mu \partial^{(x_s)}_\nu V^{bc}_{\tilde{\rho}}[x,x_s]) \tilde{j}^c_\nu[x_s]
\end{eqnarray}

\subsection{Quantization and Gauge-Fixing}

Generally, one expects the preferred field $\bar{A}^{b\mu}_c$ breaks gauge symmetry and introduces massive bosonic excitations for the fields $A^{b\mu}$.  While this is generally true, under certain cases closely connected to the presence of a dark state, a limited form of gauge symmetry can be restored.  To see this, it is helpful to bring the action found above into a more standard form by scrutinizing the change in \emph{measure} induced by removal of the detailed dynamics of the optical pumping fields.  The fluctuation of the Langevin current $\tilde{j}$ is reminiscent of the additional current \cite{Becchi74} associated with softly fixed gauge conditions - this can be seen by associating the Langevin current $\tilde{j}_\mu^a[x]\equiv \partial_\mu b^a[x]$ with the divergence of a set of scalar fields $b^a[x]$ (Nakanishi--Lautrup fields).  Integrating equation \ref{eq::Sd_v} by parts and combining with the original action,

\begin{eqnarray}
\mathcal{S}_\xi &=& \int d^4x\, [ b^a \partial_\mu(A^{a\mu}-\bar{A}_c^{a\mu}) \nonumber\\
            &-& \bar{c}^a \partial_\mu(\delta_{ac}\partial^\mu + g_A \epsilon^{abc}(A^{b\mu}-\bar{A}_c^{b\mu})) c^c  \nonumber\\
            &+&  b^a[x] \int d^4x'\, b^b[x'] \xi^{ab}[x,x'] ] \label{eq::S_ghost}
\end{eqnarray}

\noindent where we have introduced transformed ghost fields

\begin{eqnarray}
\bar{c}^a[x] = \int d^4\eta\,\, _{\bar{c}}u_i^a[\eta,x]\, \bar{c}_i[\eta] \nonumber \\
c^a[x] = \int d^4\eta\,\, _{{c}}u_i^a[\eta,x]\, {c}_i[\eta]
\end{eqnarray}

\noindent and the functions $_{\bar{c}}u_i^a$ and $_{{c}}u_i^a$ are chosen such that

\begin{eqnarray}
\kappa_{lj}[\eta]\delta(\eta-\eta') = \int d^4x \,\, _{\bar{c}}u_l^a[x,\eta] \nonumber\\
 \partial_\mu(\delta_{ac}\partial^\mu + g_A \epsilon^{abc}(A^{b\mu}[x]-\bar{A}_c^{b\mu}[x]))\,_{{c}}u_j^c[x,\eta']
\end{eqnarray}

\noindent When such a choice is possible (discussed at length below), BRST symmetry \cite{Becchi74} is satisfied, such that the action is invariant under the variation

\begin{eqnarray}
\delta \Psi &=& \delta(A^{a\mu}-\bar{A}_c^{a\mu},\bar{c}^a,c^a,b^a) =  \lambda\mathcal{Q}\Psi  \\
 & = & \lambda \, \, (\partial_\mu c^a + g_A\epsilon^{abc}(A^{b\mu}-\bar{A}_c^{b\mu})c^c,b^a,g_A\epsilon^{abc}c^bc^c,0)  \nonumber
\end{eqnarray}

\noindent which mixes bosonic fields $(b^a,A^a)$ with fermionic fields $(c^a,\bar{c}^a)$ with a scale set by an infinitesimal anti-commuting parameter $\lambda$.

Provided BRST symmetry can be maintained for the complete action, renormalization to large length and time scales may be achieved, even in the presence of the symmetry-breaking term $\bar{A}_c^{a\mu}$.  Note that the action is invariant (closed) under BRST, such that $\mathcal{Q}\mathcal{L}=0$, which can be seen from the fact that these terms are BRST-exact

\begin{eqnarray}
\mathcal{L}=\mathcal{Q}\bar{c}^a(\partial_\mu(A^{a\mu}-\bar{A}_c^{a\mu}) + \int d^4x' b^b[x'] \xi^{ab}[x,x'])
\end{eqnarray}

\noindent combined with the fact that the BRST operator is nilpotent, $\mathcal{Q}^2=0$.

The final term in the action \ref{eq::S_ghost} may be interpreted as a soft condition setting the gauge $\partial_\mu(A^{a\mu}-\bar{A}_c^{a\mu})$ to zero through the auxiliary fields $b^a[x]$, which fluctuate against a `stiffness' given by

\begin{eqnarray}
\xi^{ab}[x_1,x_2] &=& \Box_1\Box_2 V^{ab}_{\tilde{\rho}}[x_1,x_2] \nonumber \\
&=& \int d^4\eta\,\, \Box_1\bar{G}_i^a\, \kappa^{-1}_{is}\,\Box_2\bar{G}_s^b
\end{eqnarray}

\noindent with the d'Alembertian (combined with a surface term) for the $x_i$ coordinate $\Box_i=(\delta_{S_\mu}-\partial^{(i)}_\mu) \partial^{(i)\mu}$. It is important to note that this stiffness is itself a function of the $b^a[x]$ through $\tilde{j}^a_\mu$ in the functions $\bar{G}^a_i[\eta]$ and $\kappa_{is}[\eta]$, and the gauge-fixing contains higher powers than quadratic in $b^a[x]$ and its derivatives.  If the $b^a[x]$ were integrated-out of the action, this would lead to terms higher than quadratic in $A^a_\mu[x]$ and its derivatives, and would produce a contribution to the kinetic energy associated with the non-abelian field $A^a_{d\mu}[x]=A^a_\mu[x]-A^a_{c\mu}[x]$.

As a simple example, in the case $\xi^{ab}[x,x'] \rightarrow \xi_0\delta(x-x')$, with small $\xi_0$, Landau gauge conditions would be strongly enforced with a locally preferred gauge condition $\partial_\mu A^{b\mu}=\partial_{\mu} \bar{A}_c^{b\mu}$ in a more-or-less standard picture. The general case can be seen as a type of $R_\xi$-gauge fixing suitable for gauge fields produced by a natural system, in which correlation is to be expected over some length-scale, and gauge-fixing conditions are naturally softened by the dynamics of the external body.    Unlike the case for the standard model electroweak interaction, off-diagonal elements in $\xi^{ab}$ couple the $U(1)$ symmetry to $SU(2)$, and despite the fact that that the $U(1)$ sector is abelian and decoupled from the $SU(2)$ through the structure constant $\epsilon^{abc}$, all excitations can acquire mass through this term unless $\xi^{0b}=\xi^{b0}=0$ for all $b\neq0$.  Under this latter condition, the $U(1)$ fields are completely decoupled, and partial gauge symmetry may be restored, in which case a single component of the gauge field would lose its mass.

It is helpful to understand this on a more geometric footing - the gauge-fixing action represented by $\xi^{ab}$ has its origin in the action determined by products of `volume' elements $v_i[\eta]$ corresponding the coordinates $\bar{x}_i[\eta]$ - the tensor field $\xi^{ab}[x,x']$ reflects the (non-)local contributions to the total squared-volume $v_i v_j$ due to spin components $a,b$ as well as their correlations - one can view the tensor aspect as necessary to describe the local volume as due to an `internal space' representing spin dynamics and whose value reflects a contribution to the Fadeev-Popov determinant.  This is made clear by the constraint assumed above for the volume form $v_i$,

\begin{eqnarray}
\sqrt{h_i} = \bar{G}^b_i \Box b^b
\end{eqnarray}

\noindent The volume can be calculated from the coordinates $\bar{x}_i[x]$ by first using a Nambu-Goto form and introducing an auxiliary dynamical metric $_ig_{\mu\nu}[\eta,x]$ to convert it into the form of a Polyakov action

\begin{eqnarray}
v_j = \int d^4x \sqrt{g_j} g_j^{\mu\nu} \bar{h}_{\sigma\sigma'}^j \partial_\mu \bar{x}^\sigma_j \partial_\nu \bar{x}^{\sigma'}_j = \int d^4x \sqrt{g_j} h^\mu_{j\mu}
\end{eqnarray}

\noindent where $\bar{h}^{\sigma\sigma'}$ is a flat metric in the higher-dimensional Riemannian space with $\sigma=1...N$, the metric determinant $g_j=\det({_jg_{\mu\nu}})$, and $_jh^\mu_\mu$ represents the contraction of $_jg^{\mu\nu}$ with the transform metric $_jh_{\mu\nu}$ induced from the $\sigma$-coordinates as

\begin{eqnarray}
_jh_{\mu\nu} = \frac{\partial\bar{x}^\sigma_j}{\partial x^\mu} \bar{h}_{\sigma\sigma'}^j  \frac{\partial\bar{x}^{\sigma'}_j}{\partial x^\nu}
\end{eqnarray}

\noindent The constraint can be imposed by introducing a dilaton field $\phi_j[\eta, x]$ through a Lagrange multiplier $\chi_j[\eta, x]=\exp(\phi_j)$ and adding to the action

\begin{eqnarray}
S_\phi = i \int d^4x \,d^4\eta \, [ \sqrt{g_j}h^\mu_{j\mu} - \bar{G}^b_j \Box b^b ]\, \chi_j[\eta,x]
\end{eqnarray}

\noindent showing that the NL-fields $b^b[x]$ couple linearly to the exponential of the dilaton field $\phi_i[\eta,x]$.  This has an important consequence, in that the natural $R_\zeta$ gauge-fixing couples the fluctuations of the dissipative gauge field $A_d^{b\mu}[x]$ within its gauge orbit to the dilaton field.  This suggests that, within this model, the modification of measure in a path-integration associated with the Fadeev-Popov volume of the gauge orbit is tied closely to the strength of the dilaton field. This will in turn below be linked to local effects which resemble the dilaton-like components of gravitation. It is important to stress that these effects are at least partially entropic in origin,  associated with the disregard of detailed dynamics of light-scattering modes.

It is natural then to ask if the dynamical gauge field $A_d^{b\mu}$ and dynamical metric $g_{\mu\nu}^j$ obey the known equations of motion of gauge fields and gravity - for example, whether the Einstein-Hilbert form of the action is obtained in any limit.  To answer this, we return in the next section to the remaining constraints introduced by the coordinate transform above - under an appropriate redefinition of the action, these constraints may also be automatically satisfied according to the associated Euler-Lagrange equations.

Before continuing, we first interpret physically the appearance of two auxillary metrics $_jg^{\mu\nu}$ for $j=0,1$ associated to the coordinates and their conjugation.  This is related to the appearance of the volume-volume interaction appearing above. The interpretation of this term changes somewhat if we consider the combined set of coordinates as a higher-dimensional Hermitian manifold.  In this case, the dimensionality is in a sense doubled, and $v_i$ represent roots of the total volume of the manifold. Likewise, the auxillary metrics $_jg^{\mu\nu}$ resemble complex roots of the Hermitian metric, similar to vielbeins, and the inverse of the ghost-propagator appears effectively as a curvature tensor.  As such, the $_jg^{\mu\nu}$ are insufficient to fully determine the structure of the higher-dimensional manifold - with $N$ total coordinates, it remains to specify another $N$ quantities related to the spin-connections.  We will see these enter below through conditions fixing the coordinate transforms $\bar{x}$ in the form of Lagrange multipliers $\Lambda_N^\mu$.  Finally, for a Hermitian manifold, there is a natural sense of orientation to the volume form - this is provided by a final set of four multipliers $\lambda_{is}$ introduced to fix the form of the ghost propagator $\kappa_{is}$.  In this way, a single natural K\"{a}hler manifold structure is assembled through the introduction of constraints above.


\subsection{The Unruh Effect and Gauge-Field Dynamics}

The Unruh effect describes the apparent thermal occupation of a vacuum by an accelerating observer - for an observer in a frame accelerated at a rate $a$, the thermal energy scale is set by $T_u=\hbar a / 2\pi k_B c'$, with $c'$ the maximal speed of propagation of excitations of the vacuum.  Typically, Unruh effects are considered unobservably small due to the extreme accelerations necessary to achieve Kelvin-scale temperatures with excitations at the speed of light. For the experiment here, extremely low optical scattering rates are used, such that dispersive effects on the slower $\lambda$-like transition for $m_f=0,1$ lead to an effective speed of propagation \cite{RevModPhys.77.633} naively on the order of $ \Gamma c / \Omega_c^2 \sim 10^{-12}c$ at the highest bright-state Rabi-frequencies $\Omega_c$ used. For Unruh temperatures comparable to the ultra-cold thermal scales on the order of $100$nK, this yields Unruh accelerations ranging over $0.1-10m/s^2$. These are comparable to the forces generated by the optical potential, and thus should be included in the description.  To be clear, the application of the Unruh effect should not be considered surprising, since the combination of the slow-light-based calculation above and the Unruh calculation simply state that a single photon recoil of momentum is absorbed over a duration of order $1-100ms$, which is roughly the bright-state scattering rate.

Unruh effects can be considered geometric in nature, and can be captured naturally by simple geometrically-motivated scattering modes as we describe below. Thus, the dynamical gauge field is in some sense just an alternative method for describing EIT-like interference effects in a multilevel system that takes into account local variation of the optical fields.  This treatment is helpful in cases, like here, that considerable conservative motion occurs during the spontaneous photon scattering process. In the analysis above, both the interaction $V^{bc}_{\tilde{\rho}}$ and the gauge stiffness $\xi^{bc}$ above are determined by the structure of $\det\kappa_{is}[\eta]$, which in turn is limited by the Bogoliubov-Fourier wave-functions $u^k_i(\eta,x)$ through the constraints above.  It is therefore helpful that explicit $u^k_i$ be chosen.  We provide here first a somewhat incomplete description, starting from simple plane-wave structures, and then introducing arguments based on the symmetries present in the light-scattering problem constrained by conformality.  We introduce the conformal constraint in order to preserve scale-invariant dynamics already present in the action above.

The most intuitive start is to use ``plane-waves" in the new coordinates $\bar{x}[x]$ introduced above, which imply definition of the $\eta_\mu$ as conserved quantities (energy and momenta) associated with translation in the new coordinates. The complex form of the new coordinates permits smooth connection to the anomalous Bogoliubov amplitudes $u_2$ as the coordinates are varied, in line with the Unruh effect and associated analytic continuations.  For simplicity, we first propose a short form to illustrate the idea,

\begin{eqnarray}
u_l^k=\frac{\alpha_k}{2}\exp{(i\eta_\mu (\sigma^x_{tj})^s\bar{x}^\mu_j + i \pi s l )} \times \sqrt{d\mu},
\end{eqnarray}

\noindent where $s$ is summed over the values $0,1$, $\alpha_k$ is an amplitude, and $d\mu$ is a normalizing measure.  To understand this more easily, it is helpful to see these have the form

\begin{eqnarray}
u_0^k \propto \cosh(\theta) \exp{(i\phi)} \\
u_1^k \propto \sinh(\theta) \exp{(i\phi)}
\end{eqnarray}

\noindent where the Bogoliubov angle $\theta=i \eta_\mu (\bar{x}^\mu_0-\bar{x}^\mu_1)$ is given by the imaginary part of the coordinates and $\phi=\eta_\mu (\bar{x}^\mu_0+\bar{x}^\mu_1)$ by the real.

It is helpful to bring mode expansions for the ghost and scattered light modes into a similar form; for this reason we will consider below the quantity $_au_i^b = \bar{G}^b_i$, which is 'dressed' by the gauge form to transform the polarization index into one for spin, similar to the ghost mode structures $_{c,\bar{c}}u_i^b$.  Furthermore, for simplicity, we will consider the real and complex parts of the coordinates $_{A}\bar{x}^\sigma$ as a set of eight real-valued coordinates defining a higher-dimensional system, and condense the set of indices $(i,\phi=c/\bar{c}/a,b)$ with a single letter $A$.

Restricting the coordinate transforms $\bar{x}$ to be smooth and conformal, however, places surprisingly strong constraints on the form above. We will explore the complex nature in a moment - first, for purely \emph{real} coordinates, all such transformations would be members of the conformal group $SO(4,2)$. While this group includes infinitesmal generators $P_\mu$ for translation, and therefore suggests conserved momenta $\eta_\mu$ as used above, it also includes generators $M_{\mu\nu}$ for rotations and boosts to moving frames, dilations $D$, and special conformal translations $K_\mu$ related to frame acceleration. The three added generators $M,D,K$ form a subgroup isomorphic to the Weyl group $W$, such that the full conformal group is a product of an internal Weyl subspace and a Minkowski space $M_4=SO(4,2)/W$ - it is convenient to write the spatial coordinates using a matrix form

\begin{eqnarray}
X_{\alpha\alpha'}=\begin{pmatrix}
                     x_0+x_1 & x_2+ix_3 \\
                     x_2-ix_3 & x_0-x_1 \\
                   \end{pmatrix}
\end{eqnarray}

\noindent in which case all conformal transformations may be written in the form

\begin{eqnarray}
\bar{X}= \frac{AX + B}{CX + D} \label{eq::Moebius}
\end{eqnarray}

\noindent for $2\times2$ complex matrices $A,B,C,D$.

This representation is not unusual - for instance, it is used as a convenient starting point for instanton and twistor formalisms, and is simply related to the fact that all conformal mappings can be understood by manipulating a `product' of the local coordinates and the Weyl group representing an internal space. This also suggests inclusion in the above of a larger set of conserved charges $\bar{\eta}_\sigma$ associated with incorporating the irreducible representation of the internal space, and extension of the `coordinates' to a larger set $\bar{x}_{i\sigma}$ (where we still retain an index $i$ for conjugation of the complex coordinates, equivalent to a parity flip of a subset of coordinates in the higher-dimensional representation).  The plane wave expansion above may then be generalized to a pair of Fourier expansions in a basis dual (in the Peter-Weyl sense) to $SO(4,2)$ and represented by the $\bar{\eta}_{\sigma}$.

To incorporate the Unruh-like effects, the coupling between complex and real coordinates used above must be accounted for, in a way which connects the special conformal generator $K_\mu$ (representing acceleration) to the anomalous angle $\theta$. A similar approach is discussed in refs. \cite{Calixto12,Calixto11}, the first of which interestingly interprets the Unruh effect as a `spontaneous breakdown of conformal symmetry'. There, the $SO(4,2)$ group is extended into a complex representation $U(2,2)$, such that a complex (Cartan) space $D_4$ is obtained as a continuation of $M_4$.  The modified form of the conformal transformation \ref{eq::Moebius} above, irreducible representations (depending on a conformal dimension $\lambda$ and two spin indices $s_1$, $s_2$), and the transformation of an associated wave-function with the associated normalization are reported there. The complex conformal group exhibits a symmetry under the combined replacements $P_\mu \leftrightarrow K_\mu$ and $D \leftrightarrow -D$; together these imply a (CPT) symmetry under simultaneous exchange of proper time, parity, and conjugation of wave-functions \cite{Calixto11}, which the authors have argued is also tantamount to Born-Reciprocity, a point to which we will return in different language below.

The mode expansions for scattered pumping light $_au^k_i$ and (anti)-ghost modes $_{(\bar{c})c}u^b_i$ can each be expanded in a separate set of preferred coordinates as $_\phi u_i^b \propto \exp(i\bar{\eta}_\sigma \bar{x}^\sigma_{i\phi b})$, for each field $\phi=(a,a^*,c,\bar{c})$.  To bring the optical pumping modes into a form similar to the ghost modes, we define

\begin{eqnarray}
_a u_i^b = G^b_{rk} E_{op,t}^r \sigma^x_{tf} {}_au^k_g[\bar{x}_f] \sigma^x_{gs}
\end{eqnarray} 


\noindent and define all fields through coordinate transformations ${}_A\bar{x}$,

\begin{eqnarray}
_\phi u_i^b [\bar{\eta},\bar{x}] = e^{i\bar{\eta}_\sigma \bar{x}^\sigma_{A} } \label{eq::wf}
\end{eqnarray}

\noindent We include constraints that the physical momenta follow the `induced' form $\eta_{\mu} = \bar{\eta}_\sigma \partial {}_A\bar{x}^\sigma/\partial x^\mu$, and that the coordinates ${}_A\bar{x}^\sigma$ be determined by a fixed mapping ${}_A\bar{x}^\sigma[x]$ defining a four-dimensional `surface' $S$ in the higher-dimensional space.  It is formally convenient to absorb these constraints as two additional `coordinates'

\begin{eqnarray}
\bar{x}^N_A       &=& -i\log [\delta^N({}_A\bar{x}^\sigma - {}_A\bar{x}^\sigma[x])] \\
\bar{x}^{N+1}_A   &=& -i\log [\delta^4(\eta_{\mu} - \bar{\eta}_\sigma \partial {}_A\bar{x}^\sigma/\partial x^\mu)]
\end{eqnarray}

\noindent with corresponding momenta $\bar{\eta}^N=\bar{\eta}^{N+1}=1$, such that the implied sum over $\sigma$ in \ref{eq::wf} extends over $\sigma=0...N+1$, and define

\begin{eqnarray}
_\phi u_i^b [{\eta},{x}] = \int d^N\bar{\eta}\,d^N\bar{x}\,\, _\phi u_i^b [\bar{\eta},\bar{x}]
\end{eqnarray}

\noindent (note that the final coordinates $\bar{x}^{N(+1)}$ are not integrated).

The constraints above, along with the desired volume forms, reduce to a Palatini form for the action in terms of the partial derivatives ${}_Ae^\sigma_\mu=\partial{}_A\bar{x}^\sigma/\partial {x}^\mu$.  It is helpful to first break the transformations ${}_A\bar{x}^\sigma[x_\mu]$ into two steps, ${}_A\bar{x}^{j\sigma}[ x_\mu]={}_A\tilde{x}^{j\sigma}[{}_AM^j[x_\mu]]$ consisting of fixed `lifts' ${}_AM^{j}$ of $x_\mu$ to the internal ($j=0$), external ($j=1$), and constraint ($j=2$) spaces, followed by continuous mappings ${}_A\tilde{x}^{j\sigma}$ of those spaces onto themselves.  The existence of a covering group like the above allow the transformation to be broken into these three sectors.  The partial derivatives can then be written ${}_Ae^{j\sigma}_\mu={}_AM^{j\sigma'}_\mu {}_A\tilde{e}^{j\sigma}_{\sigma'}$, which allows for an interpretation of the ${}_A\tilde{e}^{j\sigma}_{\rho}=\partial{{}_A\bar{x}^{j\sigma}}/\partial{{}_A\tilde{x}^{j\rho}}$ as frame-fields or `vielbeins', through the partial derivatives of the lifts ${}_AM^{j\rho}_\mu=\partial{{}_A\tilde{x}^{j\rho}}/\partial{x^{\mu}}$.  This allows the action found above to be written in a modified Palatini form

\begin{eqnarray}
S_P = \int  d^N\bar{x} \, d^4\bar{\eta}   \,
 [ {}_A\tilde{e}^{\sigma}_{\rho}\, {}_B\tilde{e}^{\sigma'}_{\rho'} \,P^{\rho\rho'}_{\bar{\rho}\bar{\rho}'} {}_{AB}\Omega^{\bar{\rho}\bar{\rho}'}_{\sigma\sigma'} + {}_A\tilde{e}^{\sigma}_{\rho} {}_{A}F^\rho_\sigma ] \nonumber
\end{eqnarray}

\noindent where $A,B$ now represent sets $(i,j,\phi)$ condensing the labels for coordinate conjugation $i$, sector $j$ and field $\phi$.  The quantity $\Omega$ may be viewed as representing a curvature tensor, and takes its form from expansion of the ghost action above as

\begin{eqnarray}
{}_{AA'}\Omega^{{\rho}{\rho}'}_{\sigma\sigma'} = \bar{\eta}_{j\sigma} \bar{\eta}_{j'\sigma'} \bar{c}^a_i {c}^a_{i'} \delta^{\rho\rho'}\delta_{\phi,\bar{c}}\delta_{\phi',c}
\end{eqnarray}

\noindent and the contribution to the curvature from the volume constraint above is added through the action of the operator $P$ as

\begin{eqnarray}
P^{\rho\rho'}_{\bar{\rho}\bar{\rho}'} &=& \delta^A_{A'}\delta^{\rho}_{\bar{\rho}}\delta^{\rho'}_{\bar{\rho}'} + \\
 && i \chi_j \sqrt{{}_jg} {}_jg^{\rho\rho'}\bar{h}_{\bar{\rho}\bar{\rho}'} \hat{p}_{c\bar{c}\rightarrow aa} \partial_{\bar{\eta}_{j\sigma}} \partial_{\bar{\eta}_{j'\sigma'}} \partial_{\bar{c}^a_i} \partial_{ {c}^a_{i'}} \nonumber
\end{eqnarray}

\noindent which exchanges the stress-energy associated with the ghost modes with a complex curvature determined by the auxillary metric acting in the optical-pumping mode space.  Here, $\hat{p}_{c\bar{c}\rightarrow aa}$ shifts the $AB$ indices from ghost modes to optical pumping.

Classically, independent variation of the vielbeins and the spin-connections (contributing to the form of $\Omega$) for $j=0$ are well-known to lead to an action equivalent to the Einstein-Hilbert form, and thus the classical equations of motion for general relativity.  In the $j=1$ sector, the Yang-Mills kinetic energy terms arise through spin-connects in a way similar to the Kaluza-Klein compactification.  The operator $P$, if its second term is of an appropriate form, leads to a self-dual or anti-dual formulation of gravity and gauge fields.  If in this case, $P$ is to be interpreted as a type of dualization, it must be viewed as acting in the higher-dimensional manifold specified the totality of all coordinate systems $A$.

The final term $F$ in the action above goes beyond the Palatini form, arising from two sources.  The first is contributed by the ghost propagtor's covariant derivative,

\begin{eqnarray}
{}_{(\phi=\bar{c},i,j)} F^\rho_\sigma = M^{j\rho}_\mu \bar{\eta}_\sigma \epsilon^{abc}A^{b\mu}\bar{c}^a c^c
\end{eqnarray}

\noindent and represents three-leg gauge-ghost vertices, with the attachment of an antighost vielbein.

A second contribution arises from fixing the form of the ghost propagator ${\kappa}_{is}$ to the form $\bar{\kappa}_{is}$ determined above by the optical pumping fields.  To implement this, we introduce a lagrange multiplier $\lambda_{is}[\eta,x]$ and add to the action $i \lambda_{is}(\kappa_{is} - \bar{\kappa}_{is})$.  The first term is equivalent to replacing the ghost amplitudes $c_i^ac_s^a$ with $c_i^ac_s^a + i\lambda_{is}$, and the second introduces a term linear in the vierbeins to the action through

\begin{eqnarray}
{}_{A=(\phi=a,i,j)}F^{\rho}_\sigma =  -i\lambda_{is} {}_AM^{\rho}_0 \bar{\eta}_\sigma
\end{eqnarray}

\noindent In total, the action is then constructed from

\begin{eqnarray}
\mathcal{S} = \mathcal{S}_P + \mathcal{S}_\phi + \mathcal{S}_\xi' + \mathcal{S}_\psi'
\end{eqnarray}

\noindent where $\mathcal{S}_\xi'$ discards the middle term of $\mathcal{S}_\xi$ as defined above, and $\mathcal{S}_\psi'$ is second-order in the covariant derivatives of $\psi$. While it would be possible to study this model in detail (including necessities we have not included here, such as accounting for gauge-equivalence in the vielbein definitions), we will find it easier to map this problem onto boundaries of a two-dimensional space formed by consideration of the $j=2$ coordinates.

\subsection{Constraint Space as a Two-Dimensional Conformal Field Theory}

In Palatini formulations of gravity, the vielbeins $\tilde{e}$ are treated as degrees-of-freedom.  This is natural for the internal and external coordinates $j=0,1$, but merits special consideration in the constraint space $j=2$, where we have given the coordinates values defining, for $\sigma=N$, a specific coordinate transform, and, for $\sigma=N+1$, conservation of momentum.

At the outset, defining coordinates this way appears ill-posed, as only the ``point" ${}_A\bar{x}^N \rightarrow - i \infty$ carries direct meaning for the coordinate transform $_A\bar{x}$, and labels the entirety of the surface $_A\bar{x}[x]$. Likewise, only divergent values of the coordinate ${}_A\bar{x}^{N+1}$ appear \emph{a priori} relevant, representing strict momentum conservation. Though only their common divergence then would seem to have meaning, Heisenberg uncertainty suggests otherwise; fixing ${}_A\bar{x}^N$ would imply all normal coordinates are known, and thus lead to arbitrarily large fluctuations of the normal momentum.  Thus non-divergent values of ${}_A\bar{x}^N$ and ${}_A\bar{x}^{N+1}$ must also be considered in their motion.

Treating the constraint-space vielbeins on equal footing as the other two sectors requires introduction of a constraint in the form of an additional lagrange multiplier ${}_A\Lambda^{\mu}_{N(+1)}$ in the action

\begin{eqnarray}
S_\Lambda &=& i _A\Lambda^\mu_{N} (_Ae^{N}_\mu - \partial_\mu {}_A\bar{x}^N [x_\mu])  \\
S_\Lambda &=& i _A\Lambda^\mu_{N+1} (_Ae^{N+1}_\mu - \partial_\mu {}_A\bar{x}^{N+1} [x_\mu])
\end{eqnarray}

\noindent which fixes the vielbeins to the values

\begin{eqnarray}
\partial_\mu {}_A\bar{x}^N = -i {}_Ae^{\sigma}_\mu \frac{\delta_{\sigma}^N(_A\bar{x}^{\sigma} - _A\bar{x}^{\sigma}[x])}{\delta^N (_A\bar{x}^{\sigma} - _A\bar{x}^{\sigma}[x])} \\
\partial_\mu {}_A\bar{x}^{N+1} = -i \bar{\eta}_\sigma \partial_\nu {}_Ae^{\sigma}_\mu  \frac{\delta_{\nu}^4(\eta_\mu - \bar{\eta}_\sigma {}_Ae^{\sigma}_\mu )}{\delta^4(\eta_\mu - \bar{\eta}_\sigma {}_Ae^{\sigma}_\mu)}
\end{eqnarray}

\noindent It is revealing that the first of these relates the normal vector of the surface $_A\bar{x}^\sigma[x]$ to the vielbeins in all sectors, while the second relates spin-connections defined through the derivatives $\partial_\nu {}_Ae^{\sigma}_\mu$ to a four-dimensional vector tangent to the surface.  These are reminiscent of Dirichlet and Neumann boundary conditions  familiar from string-theoretical pictures, in which the gauge field is determined by the motion of the endpoint of an open string tied to the surface of a membrane. With some effort, it can be shown this is not just cosmetic.  The delta-function derivatives above may be viewed more generally as expectation values for differential operators using a common distribution function $\Delta$,

\begin{eqnarray}
\partial_\mu {}_A\bar{x}^N &=& -i {}_Ae^{\sigma}_\mu \, \langle \frac{\partial}{\partial {}_A\bar{x}^\sigma} \rangle  \\
\partial_\mu {}_A\bar{x}^{N+1} &=& -i \bar{\eta}_\sigma \partial_\nu {}_Ae^{\sigma}_\mu \, \langle \frac{\partial}{\partial {}_A\bar{p}_\nu } \rangle \\
\Delta &=& \prod_A \delta^N (_A\bar{x}^{\sigma} - _A\bar{x}^{\sigma}[x]) \times \delta^4(\eta_\mu - \bar{\eta}_\sigma {}_Ae^{\sigma}_\mu) \nonumber \\
&=&  \int D[\prod_A {}_Ap_\sigma, {}_Ax_\sigma, {}_A\bar{p}_\nu]  \delta ({}_A\bar{p}_\nu-{\eta}_\nu+\bar{\eta}_\sigma{}_Ae_\nu^\sigma) \nonumber \\
&&  \exp{\sum_A  2\pi i[{}_Ap_\sigma {}_A\bar{x}^{\sigma}+{}_Ax^\nu {}_A\bar{p}_\nu ] } \nonumber
\end{eqnarray}

\noindent where we have defined the quantity ${}_A\bar{p}_\nu$ and constrained it to the value $\bar{\eta}_\nu-\bar{\eta}_\sigma{}_Ae_\nu^\sigma$, and offset the $_A\bar{x}^{\sigma}$ by the value of $_A\bar{x}^{\sigma}[x]$.  We have chosen the names for the Lagrange multipliers $p$ and $x$, and the form of action above, to be suggestive of Born-reciprocity.

A more convenient coordinate system can be constructed by bringing the divergences of $_A\bar{x}^{N}$ to finite values through $\sigma_A=\tan^{-1}(i_A\bar{x}^{N})$.  While the endpoint $\sigma_A=\pi/2$ is maximally relevant in determining the surface location, the endpoint at $\sigma_A=-\pi/2$ is maximally irrelevant; we consider connection of this irrelevant point to another for a second surface $B$ by forming paired coordinates

\begin{eqnarray}
\sigma_{AB} = \theta(\sigma_{AB}) (\pi + \sigma_A + \sigma_B) - (\pi/2 +\sigma_A)
\end{eqnarray}

\noindent where $\theta$ is a heaviside function.  The new coordinate $\sigma_{AB}$ takes the value $-\pi$ on the surface $A$ and $\pi$ on surface $B$, and otherwise represents either of $_{A/B}\bar{x}^N$ according to its sign.  In this way, the coordinates $\sigma$ form a (Chan-Patton-like) matrix of affine parameters for curves connecting surfaces in pairs.  A similar construct $\tau_{AB}$ can be made using the $\bar{x}^{N+1}$ coordinates, taking the values $\pm\pi$ when momentum is conserved on the $A/B$ surfaces.

It is convenient to group the variables in the action also pairwise by introducing functions of the form

\begin{eqnarray}
{}_{AB}\bar{x}^\sigma[\sigma_{AB},x] = {}_{A}\bar{x}^\sigma \delta[\sigma_{AB}+\pi] + {}_{B}\bar{x}^\sigma \delta[\sigma_{AB}-\pi] \nonumber \\
{}_{AB}\bar{p}^\nu[x,\tau_{AB}] = {}_{A}\bar{p}^\nu \delta[\tau_{AB}+\pi] + {}_{B}\bar{p}^\nu \delta[\tau_{AB}-\pi] \nonumber \\
{}_{AB}{x}^\sigma[x,\tau_{AB}] = {}_{A}{x}^\sigma \delta[\tau_{AB}-\pi] + {}_{B}{x}^\sigma \delta[\tau_{AB}+\pi]  \nonumber \\
{}_{AB}{p}^\nu[\sigma_{AB},x] = {}_{A}{p}^\nu \delta[\sigma_{AB}-\pi] + {}_{B}{p}^\nu \delta[\sigma_{AB}+\pi] \nonumber
\end{eqnarray}

\noindent defined now only at the points $\sigma=\pm \pi$ and $\tau=\pm \pi$. The distribution can now be written more simply as

\begin{eqnarray}
\Delta &=&  \int D[p_\sigma, x_\mu, \bar{p}_\nu] \Delta^{-1}_p \Delta^{-1}_x \delta^4_{\tau=\pm\pi}(\bar{p}_\nu-{\eta}_\nu \mathbb{I} + \bar{\eta}_\sigma{}e_\nu^\sigma) \nonumber \\
&&  \exp{ (\frac{i}{4N_a^2} \int d\sigma d\tau \, \mathrm{Tr}[ \frac{d}{d\sigma}(p_\sigma \bar{x}^{\sigma}) + \frac{d}{d\tau} (x^\nu \bar{p}_\nu) ] )} \nonumber
\end{eqnarray}

\noindent where we have suppressed the $A/B$ matrix indices, $\mathbb{I}$ is the matrix identity in the $A/B$ space, $N_a$ its dimension, and the trace is taken over the matrix products of $p_\sigma \bar{x}^{\sigma}$ and $x^\nu \bar{p}_\nu$.  Note also that we have taken advantage of the form of XX as a full derivative to introduce integration over $\tau$ and $\sigma$.  The factors of $\Delta_p =\int D[p_\sigma]$ account for the increase in measure of the functional integral by introducing interior points $p_\sigma[\sigma]$ for $-\pi < \sigma < \pi$, which do not affect the integrand.  All such functions must by definition be antisymmetric in $\sigma$ under transposition of the Chan-Patton indices.  Finally, note that the partial derivative $e_\nu^\sigma$ (only presently defined at the values $\tau=\pm\pi$) has also been extended to a matrix form - this implies generally noncommutative relations $[e^\sigma_\mu,e^{\sigma'}_{\mu'}]\neq 0$, and becomes relevant if the same form is also used to represent the Palatini action.

Since the surface normals form an $N-4$ dimensional subspace, it is simpler to group the multipliers $\pi_\sigma=(x_\mu, p_{\sigma'})$ into an $N-$dimensional vector, and form a corresponding vector $q_\sigma=({\eta}_\nu \mathbb{I} - \bar{\eta}_\sigma{}e_\nu^\sigma,\bar{x}_{\sigma'})$.  In these variables, the distribution can be written concisely as

\begin{eqnarray}
\Delta &=&  \int D[\pi_\sigma, \tilde{g}]\, \Delta_g  \\
&& \exp{  \int \, \mathrm{Tr}[ \frac{i}{4N_a^2} d(\pi_\sigma q^{\sigma}) - \frac{1}{2}\chi\star (d\pi^\sigma) \wedge d\pi_\sigma ] } \nonumber
\end{eqnarray}

\noindent with $d$ the exterior derivative on the $(\sigma,\tau)$ space, $\star$ the hodge dual, and $\wedge$ the exterior product.  Note that integration is now also taken over the two-dimensional metric $\tilde{g}$, which in combination with the final term of the exponent, accounts for the measure of integration over interior points; we will assume $\det{(\tilde{g})}=1$, and introduce Weyl-like scalings through the dilaton field $\chi$, which must coincide with that introduced above to match the boundary values.  The Fadeev-Popov factor $\Delta_g$ appears due to gauge symmetry in $\tilde{g}$.  Aside from the matrix-nature of the variables, the form for $\Delta$ above is equivalent to that described in \cite{Leigh13} for a bosonic string. There, it was noted that completing integration over $\pi^\sigma$ results in the standard Polyakov action with coordinates $q^\sigma$ and an inverted coupling constant $\chi$, while introducing an integration over $q^\sigma$ results in a Polyakov action in $\pi^\sigma$ - a realization of T-duality.  Therefore, perhaps not surprisingly, there is motivation for use of these ideas without \emph{a priori} postulation of strings, but only beginning from a description of gauge field dynamics. The advantage is that non-perturbative techniques in string theory, such as T- and S-duality, are now directly applicable to these problems; it is also convenient that two-dimensional conformal field theory may be applied to the fields $\bar{x}^\sigma$ defined on the $(\sigma,\tau)$ coordinate space.

\subsection{Holography and Effects on Excitation Masses}

The picture of T-duality is particularly useful in understanding how dark-state conditions can lead to a discrete (topological) change in non-equilibrium behavior.  Generally, the energy-momentum relation of excitations of a string are given discretely by its winding numbers $m$ along compact dimensions, and its integral $n$ number of phase windings in the fourier-decomposition of ${}_{AB}\bar{x}^\sigma$ along its length $\sigma$.  T-duality asserts (for the bosonic string) that the same energy-momentum relation will be found on combined exchange $m \leftrightarrow n$, exchange of the $\bar{q}^\sigma$ fields for a dual set, and a specific inversion of the coupling-constant $\chi$ and length of the compact dimensions.  Finally, for open strings as above, Neumann and Dirichlet boundary conditions are exchanged. The field $\pi^\sigma$ above can be seen as the field dual to $\bar{q}^\sigma$, and thus if it experiences a change which would alter these discrete numbers, the excitation spectrum is changed.

The coordinate transform introduced for normal-mode scattering of optical pumping light above introduces an extra factor in the constraint sector for the coordinate $_A\bar{x}^{N}$ related to the amplitude $O_a$ of the prefactors $G_{rk}$ and $E_{op}^r$.  The corresponding coordinate ${}_A\bar{x}^N$ with $A=(\phi=a,i)$ therefore enters everywhere offset by the value $i\log O_a$, making it convenient to absorb the offset into the constraint-space description.  Since the effect enters only through the expectation values against the distribution $\Delta$, it is clear this is equivalent to a local shift of the value of the multiplier

\begin{eqnarray}
\pi_\sigma[\sigma,\tau] \rightarrow \pi_\sigma + \frac{i}{O_a}\frac{\partial{O_a}}{\partial{x^\mu}}e^\mu_\sigma
\end{eqnarray}

\noindent which is tantamount to a gauge field derived from the extension of $O_a[x^\mu[\bar{x}^{\sigma'}]]$

\begin{eqnarray}
\alpha_{\sigma\beta} = \partial_\beta \frac{i}{O_a}\frac{\partial{O_a}}{\partial{x^\mu}} e^\mu_\sigma[\sigma,\tau]
\end{eqnarray}

\noindent introduced into the two-dimensional conformal field theory for the fields $\pi^\sigma[\sigma,\tau]$ ($\beta$ indexes the 2D coordinates $(\sigma,\tau)$).  Since the only physically-relevant points for these expectation values lie on the boundary at $\sigma,\tau=\pm \pi$, it is natural to expect the impact of this term can be understood through a line-integral of this quantity tracing the boundary of the constraint space.  This is apparent in XX above, as the first term in the `action' defining $\Delta$ is a boundary term.  Due to the shift of $\pi^\sigma$, this contributes $S_h = - \sum_\sigma {i h^\sigma}/{4 N_a^2}$, where


\begin{eqnarray}
h^\sigma =  \oint_{\partial\Sigma} \alpha^\sigma_\beta \cdot d\ell^\beta
\end{eqnarray}

\noindent Through the definition of ${q}^\sigma$, we can see that the integral over the boundary $\partial \Sigma$ of the constraint space $\Sigma$ is equivalent to some loop $C_{\bar{x}}$ in the full coordinate systems $_{A}\bar{x}^\sigma$ demarcating the boundary of the two-dimensional system, and lying within one (or more if they intersect) of the surfaces $A$.  Since these are themselves parameterized by $x^\mu$, the loop can be projected into the original coordinates as $C_x[x]$. Equivalently then, the contribution from the constraint space to the full action is given by integrals of a gauge-like quantity related to the dynamical gauge-field - these may be viewed equivalently as (Wilson-like) loops in real space, or loops along the boundary of a two-dimensional field theory. The geometry of the loops arise dynamically according to the variation of the vielbeins $\tilde{e}$, and the field structure is defined by the arrangement of optical pumping fields and external magnetic fields.  Finally, this quantity is insensitive to the internal parameterization of the constraint space, and is therefore independent of the gauge symmetry inherent in $\tilde{g}_{\alpha\beta}$.

Alternatively, the effect of these loops can be understood from the \emph{interior} of the constraint space $\sigma,\tau$ using the two-dimensional field curvature $f^\sigma=\epsilon_{\alpha\beta}\partial_\alpha \alpha^\sigma_\beta$.  In this way, the discrete values of $h^\sigma$ can be seen to represent flux quanta of the gauge-field $\alpha^\sigma_\beta$.  At a base level, the Virasoro algebra representing the two-dimensional field theory is altered by the introduction of additional terms in the stress-energy tensor $T(z)$ due to topological defects in $\alpha^\sigma_\beta$; as a result, the generators $L_n$ are shifted, and the ground state $|0\rangle$ modified such that $L_0 |0\rangle \neq 0$.

Not all flux of $\alpha^\sigma_\beta$ are relevant to the gauge-boson dynamics. The excitation modes of a string corresponding to gauge-field dynamics are given by vertex operators of the fields $q^\sigma$ \cite{Leigh13} as

\begin{eqnarray}
V \sim \int_{\partial\Sigma} ds \, \zeta_a : \partial_z \bar{q}^a \exp{(i \eta^\gamma_\sigma \bar{q}^\sigma )}:
\end{eqnarray}

\noindent where $\zeta_a$ is a polarization vector, the sum over index $a$ is restricted to directions parallel to the surface, the derivative $\partial_z$ is taken with respect to $z=\exp{(i\sigma - \tau)}$, and the $::$ imply normal ordering in $\tau$.  While the constraint of Weyl-invariance normally restricts gauge-bosons to zero-mass $\eta^\gamma_\sigma \eta^{\gamma\sigma}=0$ \cite{Tong}, the presence of an offset to the dual field $\pi^\sigma$ with non-vanishing field curvature implies a shift to the corresponding $n$ and $m$, generally lifting this condition and giving the gauge boson mass.  If, through manipulation of $O_a$, this defect is removed, the massless condition is restored.  A similar statement may be made concerning conditions on transversivity of polarization $\zeta^a \eta^\gamma_a=0$, which is, in the absence of the offset, a result of requiring the operator above to be primary.  An excellent viewpoint for this was suggested by ref. \cite{Leigh13}, in which a T-dual vertex was introduced, equivalent here to

\begin{eqnarray}
V^\star \sim \int_{\partial\Sigma} ds \, \zeta_b : \partial_z {\pi}^b \exp{(i\eta^\gamma_\sigma \oint_{z_i} \star d\pi^\sigma )}:
\end{eqnarray}

\noindent The effect of a puncture introduced by the shift of $\pi_q$ at $z_i$ is then to introduce an associated winding number about $z_i$.

In this way, we see that the holographic principle, which associates a particular scaling associated with the `unit area' defined by a Wilson loop, and the vanishing of a gauge-boson mass gap are related in the current problem through the experimental `knob' offered by $O_a$.  It is worth mentioning that a similar mechanism is known to occur in supersymmetric Yang-Mills and instanton \cite{TongInstanton} pictures, in which the Higgs mechanism is associated with the separation of otherwise coincident D-branes.  In these cases, the potential energy of the surfaces corresponding to ${}_{AB}\bar{x}^\sigma$ is restricted by T-duality and supersymmetry, and proportional to $[\bar{x}^\sigma,\bar{x}^{\sigma'}]$ for directions normal to the surface.  Within the zero-energy subspace obtained by choosing commuting coordinates, the gauge boson mass is proportional to the surface separations (or in the case of instanton physics, the separation of the instanton and D-brane).  Though it is likely a similar scenario may be derived here (by integrating out the multipliers $_{A}\Lambda_N^\mu$), we will not pursue it further.

\subsection{Disorder and Thermal Averaging}

It is helpful to put holographic effects on a more physical footing before considering the quantitative details entailed by the model above.  We consider first the boundary separating any given region of an atomic wave-packet (near which its dynamics are related by unitary motion) from regions in which only thermal correlations are present, which occurs roughly at the thermal deBroglie wavelength.  The slowest dynamics of this system, those measured in the experiment, occur at low temperatures through the continuous motion of this boundary. In a disordered system, increasing the radius $r_{dB}$ of this boundary changes free energy associated with the unitary region in two distinct ways.  Trivially, as the packet cools, the thermal entropy decreases, and the wavepacket expands. In a disordered system, however, the informational entropy of disorder also increases as a larger region of disorder is sampled, and thus there is a balance required between these rates of entropy change. A non-trivial outgrowth of the above is that these two effects interact with a third effect associated with the change in a local analog gravitational potential through the dilaton-field.

Thermal averaging can be understood in the usual way by adding a complex `time' coordinate $t_t$ to the action, and assuming all dynamical fields are periodic in the new direction over a scale $\beta_t$ with $1/\beta_t = k_B T$, and $T$ the temperature. Likewise, to understand the role of a disordered optical field formed by the far-detuned light, we introduce an additional parameter, or coordinate, $t_d$ to those describing the mode amplitudes $a_{fi}[x_\mu,t_t,t_d]$ for far-detuned light, as well as all other dynamical variables.  We will assume this coordinate also to imply periodicity, though at an independent inverse `temperature of disorder' $\beta_d$.  Corresponding to each are additions to the original action proportional to $|a_{fi}|^2$ and the associated kinetic energy. Physically, the disorder temperature is related to the variance $\delta a_{fi}^2$ of the far-detuned modes by $k_B T_d = \delta a_{fi}^2$.  Though one could follow similar steps to those for integrating-out the optical pumping modes above, we will forgo that complication, and concentrate instead on changes to the constraint space.

Since we assume no particular relation to the prior set of preferred coordinates ${}_A\bar{x}^{j\sigma}$ for $j=0,1,2$, it is sufficient to consider the additional coordinates $(t_t,t_d)$ as an isolated sector $j\equiv -1$ covered by the other coordinates.  In this way, the entirety of the previous analysis may be retained, with the exception of expanding the indices to include the larger range of $j$, and with coordinate integrations in the action extending over all $j \le 1$.  We therefore anticipate the addition of a new set of vielbeins $\tilde{e}^{j=-1,\sigma}_\rho$ to our dynamical variables.  We consider here the change to the \emph{constraint space} dynamics due to the presence of the additional associated partial derivatives ${e}^{j=-1,\sigma}_\mu$ and constraint-space fields ${q}^{j=-1,\sigma}$.

An immediate consequence can be seen from T-dualities, under which the coordinate lengthscales $\beta_t$ and $\beta_d$ independently map onto dual pictures in which the corresponding Polyakov actions appear with coupling constant and dimension related by $\beta_i \rightarrow \chi / \beta_i$ with $i=t$ or $d$.  Applying the T-dual (integrating out ${q}^{j=-1,\sigma=t}$) to one (i=t) of the two shows that the action for any particular string state is altered by $Z_t \chi / \beta_t$, where the integer $Z_t$ is determined by the defect(s) introduced by $O_a$ above. In addition, the Dirichlet and Neumann boundary conditions are exchanged, leading to a total of $N+2$ fields corresponding to each.

Without modifying the action, another type of exchange can be made to the same dimension $t_t$ exploiting modular invariance to exchange coordinates $\sigma$ and $\tau$ in the argument of $\pi^\sigma[\sigma,\tau]$.  Similar arguments, using modular invariance alone, lead to the Cardy-Verlinde relation\cite{Cardy,VerlindeCardy} for the entropy of a single field component $q^\sigma$ or $\pi^\sigma$ as $S^\sigma(L^\sigma_0)=2\pi\sqrt{(c^\sigma/6)(L^\sigma_0-c^\sigma/24)}$ with $c^\sigma$ the central charge for the field with index $\sigma$, and $L_0^\sigma$ the virasoro generator representing the ground state. The combined operations allow the highly-disordered and low-temperature actions to be mapped onto one another, in that the distribution $\Delta[\beta_d,\beta_t]=\Delta[\beta_d,\chi/\beta_t]$.  As a result, the action for $\Delta$ is minimized (that is, the entropy is maximized under exchange of energy within the $j=-1$ sector with all other variables held fixed) when the dilaton field assumes the value $\chi=\beta_t\beta_d$.  This suggests the `local strength of gravity' is affected by both the strength of disorder and the thermal energy scale.  Since the Planck-area is defined using the gravitational constant and effective speed-of-light, this suggests that holographic scales will be set by the same physics.

\subsection{Optical Fiber Properties}
\label{ofp}

We model the optical properties of the fiber using all solutions of Maxwell's equations in a cylindrically-symmetric dielectric waveguide~\cite{lpModes} described by electric field strengths in cylindrical coordinates,

\begin{equation}
\begin{split}
& \vec{E}_{\nu_m,K_m} =\\
 & \left [ \vphantom{\frac12} \frac{i}{K_m^2}\left(\beta_m K_m A J_{\nu_m}' (K_m r)+ i\nu_m \omega \mu \frac{B}{r} J_{\nu_m}(K_m r)\right)\hat{\mathbf{r}} +  \right. \\
 & \frac{i}{K_m^2}\left(\beta_m i \nu_m \frac{A}{r} J_{\nu_m} (K_m r) - B\omega\mu K_m J_{\nu_m}' (K_m r)\right)\hat{\bm{\phi }} + \\
 &  \left. A J_{\nu_m} (K_m r)\hat{\mathbf{z }}  \vphantom{\frac12} \right ] e^{i(\nu_m \phi +\beta_m z)}
 \label{Eq:modesCylindrical}
\end{split}
\end{equation}
where $m$ indexes all solutions to the characteristic equation, $\nu_m$ is the (integer) azimuthal mode index, $K_m$ the radial wave-vector, and $\beta_m$ the longitudinal wave-vector. $J_\nu$ are $\nu^{\text{th}}$-order Bessel functions of the first kind, and the constants $A,B$ are provided in reference~\cite{lpModes}.  Through the course of propagation, an initially well-defined superposition of modes becomes, through random intermodal coupling, a statistical distribution of occupied modes with amplitudes we designate by $\{a_{m}\}$.  We argue that the most statistically relevant superposition is defined by a maximum entropy argument to distribute power evenly between modes in the fiber.  We represent the collapse of information contained in the many mode amplitudes $a_{m}$ into a typical region of space where beams interfere by constructing a local power-series representation of the electric field from all four fibers. The power series is computed with some effort by first expanding the electric field for a given beam to all orders, $\sigma$, in cylindrical radius $r_b$ about its own optical axis,

\begin{equation}
\begin{split}
\vec{E}_f &= \sum_{m,b}  a_{m,b} \hat{E}_{\nu_m,K_m, b} \\
&= \sum_{m , b} a_{m , b} \sum_{\sigma, k_b} (K_{m} r_b)^\sigma \mathcal{E}_{m ,\sigma, k_b} (\phi_b, z_b) e^{i(\nu_m \phi +\beta_m z)} \hat{k}_b
\end{split}
\end{equation}

\noindent where $b$ indexes the beam, $\hat{k}_b \in (\hat{r}_b,\hat{\phi}_b,\hat{z}_b )$ are cylindrical unit vectors defined along each beam axis, and $r_b,\phi_b,z_b$ are the corresponding coordinates. Combining all beams into a common cartesian coordinate system $(x,y,z)$ centered at the beam intersections and scaled by a variable length $\ell$, and expanding the result in a local power series near the origin as $\vec{E}_f = E_{k,l_x,l_y,l_z}  (2x/\ell)^{l_x} (2y/\ell)^{l_y} (2z/\ell)^{l_z} \hat{k}$, the coefficients $E_{k,l_x,l_y,l_z}$ may be determined from the mode amplitudes $\{a_{m,b}\}$ through a matrix $\Lambda$ as $E_i = \Lambda_{ij}a_j$. This expansion was checked by confirming that $\vec{E}_{\nu_m,K_m}  - E_{k,l_x,l_y,l_z}  x^{l_x} y^{l_y} z^{l_z} \hat{k} = 0$ to machine precision within a radius of 1 micron from the origin.

\begin{figure}[htp]
\includegraphics[width=3.25 in]{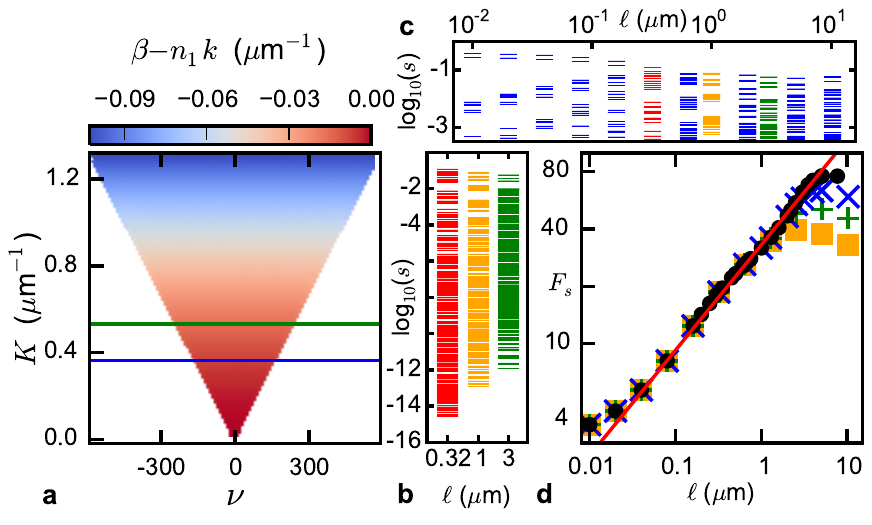}
\caption{
\textbf{Characterization of the optical modes that form the disordered potential.}
\textbf{a}, Solutions of the characteristic equation for the optical fiber for all possible modes denoted by $\nu$, $K$ pairs.
The color scale denotes the deviation $\beta - n_1 k$, with $n_1$ the core index and $k=2\pi/\lambda$, for wavelength $\lambda$, of the propagation constant $\beta$ from that of an axial ray.
The green (blue) horizontal line indicates the lower (upper) boundary of the occupied modes for the only high-$K$ (low-$K$) data in Fig.~\ref{holding} \textbf{c} and Fig.~\ref{fig::differentScalings} \textbf{e,f,g} (\textbf{c,d,g}).
\textbf{b}, The information content projected into a volume of dimension $\ell$ in the experiment chamber is apparent from the contraction of normalized singular values of the expansion matrix $\Lambda$ as the scaling parameter $\ell$ is increased.  All singular values of $\Lambda$ constructed to 10th order in the power series expansion are shown for $\ell=0.32 \mu$m (red), $\ell=1.0 \mu$m (orange), and $\ell=3.0\mu$m(green).
\textbf{c}, Most significant singular values for all length-scales $\ell$.  The diffraction limit is visible as a closing of the gap between the largest three eigenvalues as the length-scale is increased to $O(1 \mu \text{m})$.
 \textbf{d}, Scaling of information content within a volume as determined by the eigenmode structure of the fiber, and mode projections into experiment chamber, demonstrating $F_s \sim \ell^\kappa$, with $\kappa=0.54\pm0.06$.  The power series shown contain all terms up to order 6 (squares), 8 (pluses), 10 (crosses), and 12 (circles).  The red line shows the best fit between $0.04\mu$m and $3.6\mu$m for 12$^{\text{th}}$ order, which is used to extract $\kappa$.  Its uncertainty corresponds to the standard deviation of pairwise slopes for all points presented here.
}\label{fig::mode_scaling}
\end{figure}

We characterize the effective number of degrees of freedom, $F_s$, in the electric field by taking the exponential of the Shannon entropy of $\Lambda$,

\begin{equation}
F_s = e^{ - \sum_{i} s_i \ln s_i }
\end{equation}

\noindent where $s_i$ are the the singular values $\sigma_i$ of $\Lambda$ normalized to their sum, $s_i=\sigma_i/\sum_i \sigma_i$.  A plot of the singular values $s_i$ is shown in Fig.~\ref{fig::mode_scaling}b and c, illustrating for distances short compared to the optical wavelength three effective degrees of freedom for the three directions of electric field, for longer distances a sub-extensive power-law scaling due to long-range correlations in the field strength, and at larger distances a breakdown of the power-series expansion, which occurs later for a larger number of terms in the expansion.  The value of $\kappa$ is extracted from a best fit over the scaling region for a maximum term in the expansion of $l_x+l_y+l_z=12$, which consists of approximately two decades in length scale.

To vary the information scaling law described by $\kappa$ and thus interrogate the holographic scaling relation described in the main text, non-equilibrium mode distributions were created in the fiber by varying the launch conditions.  While equilibrated mode content was created using a strong lens to focus light near the entrance facet of the fiber at a numerical aperture just below that of the fiber, under-filling of modes was accomplished by utilizing the comparatively slow redistribution of mode content between modes with large differences in $K_m$. Non-equilibrated conditions were generated by changing the angle and inclination of the injection of the beam to the fiber axis. $\kappa$ from each of five different under-filled experiments are plotted in the inset of Fig.~\ref{holding}c. Two types of non-equilibrated conditions were generated by either (I) injecting a collimated beam along the fiber axis which predominantly populates the low-radial modes or (II) injecting a focused beam with an inclination to the fiber axis which  populates the high-radial modes.

$\kappa$ was calculated for each of these conditions using the method described above but with $\Lambda$ multiplied by an additional mask that sets all high-radial (low-radial) modes greater (less) than $\text{max}(K_m) \times K_\text{rad} / K_\text{M}$  to zero. $K_\text{rad}$ is equal to the radius of the beam in the Fourier plane determined by measuring where the average radial intensity falls to 50$\%$  of the intensity at the center, and $K_\text{M}$ is the value of $K_\text{rad}$ with equilibrated mode content. To verify the robustness of the calculation of $\kappa$, we also weighted each of the modes of $\Lambda$ by either a gaussian with an envelope whose standard deviation is equal to $\text{max}(K_m) \times K_\text{rad} / K_\text{M}$, or to a seventh degree polynomial spline fitted to the radially averaged distribution of intensities of the beam in the Fourier plane. The gaussian weighting yielded $\kappa$ to within $\pm 0.02$, and the spline yielded $\kappa$ to $\pm 0.03$.

\begin{figure*}[htp]
\includegraphics[trim={0 0.5in 0 20cm},clip,width=7in]{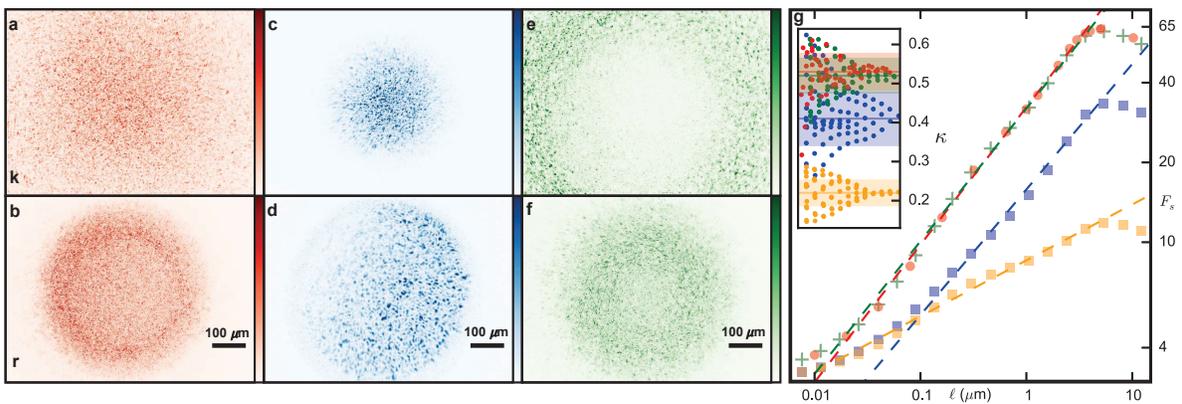}
  \caption{\textbf{Analysis of information content of optical modes.}
  Images of the intensity distribution in the aperture (\textbf{a,c,e}) and imaging (\textbf{b,d,f}) planes.
   Red, green, and blue  correspond  to all, only low-$K$, and only high-$K$ optical modes from the light potential, respectively.
  \textbf{g}, Scaling laws of the information content of the corresponding figures in \textbf{a-f}.
  The exponential of the Shannon information, $F_s$, is calculated for each point as a function of distance from the origin, $\ell/2$, constructed with all terms up to 10th order in the power series.
  The lines are least-squares linear fits to log$(F_s)$ for $\ell$ between $40$ nm and $3.6 \mu$m, which yield exponents, $\kappa$, of $0.53\pm0.05$ (red), $0.52\pm0.05$ (green), $0.41\pm0.07$ (blue).
  All possible $\kappa$ from each combination of two points of log$(F_s)$ are shown in the inset; the error on the fit was taken to be the standard deviation of this spread in slopes.
  Orange is a simulation with only  the lowest 5\% modes in $K$ retained with $\kappa = 0.22\pm 0.03$. See text (Methods \ref{ofp}) for more details.
}
\label{fig::differentScalings}
\end{figure*}

\subsection{Microwave Spectroscopy and Vibrational Spectroscopy Measurements}
We performed a microwave spectroscopy measurement to verify the cooling effect in the disordered potential with a wide range of vibrational frequency distribution. Atoms are first cooled for one second under optimized conditions, with the optical pumping beam 12 MHz blue detuned to the $|F=1\rangle \to|F=0\rangle$ transition on the $D_2$ line. In addition, another weak beam resonant with the $|F=2\rangle \to |F'=2\rangle$ transition on the $D_2$ line is applied to depump atoms from $|F=2\rangle$ hyperfine manifold. At the end of the cooling period, the optical pumping and depumping beams are turned off, while the magnetic field remains unchanged. Shortly (25 ms) after the optical pumping beam is off, a microwave field is applied with varied frequency for $50$ms, together with weak (``blow-off") light resonant with the $F=2\to F'=3$ transition to remove atoms from the potential. Negligible loss of atoms is observed without application of  microwave radiation, indicating atoms are well de-pumped during cooling to the $F=1$ manifold. The microwave frequency is tuned to $f=f_0+\delta f$, where $f_0$ is the hyperfine splitting, and $\delta f$ is a frequency offset chosen in the range $\pm$150 kHz, causing transfer of atoms to $F'=2$ and subsequent loss due to resonant light scattering. The remaining atoms are recaptured in a MOT to measure the atom number through fluorescence imaging.

As shown in Fig. \ref{rscooling}a in the main text, loss occurs at evenly spaced peaks corresponding to microwave coupling on allowed transitions between different states in the $F=1,2$ hyperfine manifolds. The separation is measured to be 40.5 kHz, corresponding to a Zeeman shift from external magnetic field of 57.8 mG.  The slight asymmetry in peak shape shows a cooling signature. Though hard to define an equilibrium temperature for atoms, we extract a kinetic temperature by fitting the peak shape with a Boltzmann weighted Lorentzian as

\begin{align}
N&=N_0 -A\sum_{n_1,n_2} C_{n_1,n_2} \int \text{d}\nu   \nonumber \\ & P(\nu) e^{-h\nu n_1/k_B T}(1-e^{-h\nu/k_BT})\frac{1}{1+4\delta_f^2/\gamma^2}
\end{align}

\noindent where $P(\nu)$ is the measured vibration frequency distribution (described below), $\gamma$ is dominated by broadening from the ``blow-off" light, and the detuning $\delta_f = f - (n_2-n_1) \nu -\delta_z$ with $\delta_z = g \mu_b\Delta m_F$ the Zeeman shift, with $g \mu_b = 700$kHz/G. Here, $C_{n_1,n_2}$ is the coupling coefficient between two different vibrational states
\begin{align}
C_{n',n}=\sum_{l=0}^{\text{min}(n',n)}\frac{\sqrt{n!~n'!}\,e^{-\frac{1}{2}\alpha^2}(-\alpha)^{(n'-l)} \alpha^{(n-l)}}{(n-l)!~(n'-l)!~l!}
\end{align}
where $\alpha=\sqrt{m\omega \delta x^2 /2\hbar}=\delta x/\sqrt{2}x_0$, with $x_0$ the harmonic oscillator length scale. $\delta x$ denotes a position shift of potential minimum for two different spin states. The temperature is estimated to be 50 nk.	

To measure the vibrational frequency distribution of atoms in the disordered potential, we parametrically excite atoms by modulating the laser intensity with small amplitude. For far off resonance dipole trap, the heating rate $\Gamma$ is given by \cite{PhysRevA.56.R1095},
\begin{equation}
\Gamma=\pi^2 \nu^2 S(2\nu).
\end{equation}
Here $\nu$ is the local trap frequency and $S(\omega)$ is the one-sided power spectrum of the fractional intensity noise. Due to the even symmetry of the parametric modulation, the heating rate depends on twice the trap frequency. The average energy of atoms in the modulation bandwidth increase exponentially, $\langle E(t)\rangle=\langle E_0\rangle\exp(\Gamma t)$.

The measurement start by cooling atoms for 2.1s at optimum condition.  After that, we modulate the speckle beam intensity for 100 ms with a chirped sinusoidal form
\begin{equation}
\Delta I/I= A \sin ((2\pi (f_A-0.5-t/T)t)
\end{equation}
where  $A$ is the modulation depth, $f_A$ is the modulation center frequency and $T$ is the modulation time. The modulation depth $A$ is set to be 2\% to minimize the power broaden effect. The chirp frequency bandwidth is 1 kHz.  Due to the heating of this modulation, atoms possessing vibrational frequencies within this modulation bandwidth are heated and lost from the potential when their energy are sufficiently higher than the potential depth. After that, atoms are holding in the disordered potential for extra 100 ms and then collected in a MOT to measure the atom number. The result is shown in Fig. \ref{rscooling}b in the main text, as the modulation frequency $f_A$ is scanned. The plot is average over 591 measurements, with each frequency chosen randomly in the range of 1 to 125 kHz to minimize effect from atom number drift over time. We further extract a normalized vibrational frequency distribution by taking account the modulation power spectrum,
\begin{equation}
N=N_0-N_0\int P(\nu)H(e^{\pi^2\nu^2 S_M(2\nu)t},\langle E_{th}\rangle/\langle E_0\rangle)\text{d}\nu ,
\end{equation}

\noindent where $H$ is the Heaviside step function , $\langle E_{th}\rangle$ is a threshold energy, and $S_M(\omega)$ is the modulation power spectrum. The vibrational frequency distribution $P(\nu)$ is shown in the inset of Fig. \ref{rscooling}b in the main text.

\bibliography{references}
\end{document}